\documentstyle[prb,aps,epsfig,a4]{revtex}
\begin{document}
\title{\bf Study of the $\pi\pi$ mass spectra \\ in the
           process  $e^+e^- \to \pi^+\pi^-\pi^0$ at
	   $\sqrt[]{s} \simeq 1020$ MeV.}
\author{ M.N.Achasov\thanks{ E-mail: achasov@inp.nsk.su,
FAX: +7(383-2)34-21-63},
         V.M.Aulchenko, K.I.Beloborodov, A.V.Berdyugin,A.G.Bogdanchikov,
	 A.V.Bozhenok, A.D.Bukin, D.A.Bukin, S.V.Burdin, T.V.Dimova,
	 V.P.Druzhinin, V.B.Golubev, V.N.Ivanchenko, P.M.Ivanov, A.A.Korol,
	 M.S.Korostelev, S.V.Koshuba, A.V.Otboev, E.V.Pakhtusova,
	 E.A.Perevedentsev, A.A.Salnikov, S.I.Serednyakov,  V.V.Shary,
	 Yu.M.Shatunov, V.A.Sidorov, Z.K.Silagadze, A.V.Vasiljev, 
	 Yu.S.Velikzhanin }
\address{
          Budker Institute of Nuclear Physics,  \\
          Siberian Branch of the Russian Academy of Sciences and \\
	  Novosibirsk State University, \\
	  11 Laurentyev,Novosibirsk, \\
	  630090, Russia}
\maketitle

\begin{abstract}
 The invariant mass spectra of the $\pi^+\pi^-$ and $\pi^\pm\pi^0$ pairs
 in the process $e^+e^- \to \pi^+\pi^-\pi^0$ were studied in the SND
 experiment at the VEPP-2M collider in the energy region
 $\sqrt[]{s} \simeq 1020$ MeV. These studies were based on about
 $0.5 \times 10^6$ experimental events. The spectra were analyzed in the
 framework of the vector meson dominance model. It was found that the
 experimental data can be described with
 $e^+e^- \to \rho\pi \to \pi^+\pi^-\pi^0$ transition only. Upper limit
 on the branching ratio of the $\phi(1020)\to\pi^+\pi^-\pi^0$ decay through
 intermediate states different from  $\rho\pi$ was obtained at the 90 \%
 confidence level: $B(\phi\to\pi^+\pi^-\pi^0)<6 \cdot 10^{-4}$. 
 The $\rho$-meson mass and width which follow from the spectra analysis are
 $m_\rho=775.0\pm 1.3$ MeV, $\Gamma_\rho=150.4 \pm 3.0$ MeV. Neutral and
 charged $\rho$-mesons mass difference was found to equal
 $m_{\rho^\pm}-m_{\rho^0}=-1.3\pm2.3$ MeV. In the $\pi^+\pi^-$
 mass spectrum the $\rho-\omega$ interference was seen at
 two standard deviations level.
\end{abstract}

\section{Introduction}

 In the framework of the vector meson dominance  model (VDM) the cross section
 of the process $e^+e^- \to \pi^+\pi^-\pi^0$ in the energy region
 $\sqrt[]{s} \sim 1$ GeV is determined by the amplitudes of vector mesons V
 ($V=\omega,\phi,\omega^\prime,\omega^{\prime\prime}$) transitions into the
 final state: $V \to\pi^+\pi^-\pi^0$. From experimental data it is known
 that the $\rho\pi$ intermediate state dominates in these transitions. 
 In the energy region $\sqrt[]{s} \simeq m_\phi$ the main contribution to
 the cross section comes from the $\phi(1020)$-meson decay
 $\phi\to\rho\pi\to\pi^+\pi^-\pi^0$. The total cross section of the process
 $e^+e^-\to\pi^+\pi^-\pi^0$ in the vicinity of the $\phi(1020)$ resonance
 was studied in several experiments \cite{orse1,olia,orse2,nd,cmd1,cmd2},
 including the SND (Spherical Neutral Detector) experiment at VEPP-2M collider
 \cite{phi98}. Studies of this process dynamics 
 (the dipion mass distribution) were reported only in two works
 \cite{cmd2,orse3}.

\begin{figure}
\begin{center}
\epsfig{figure=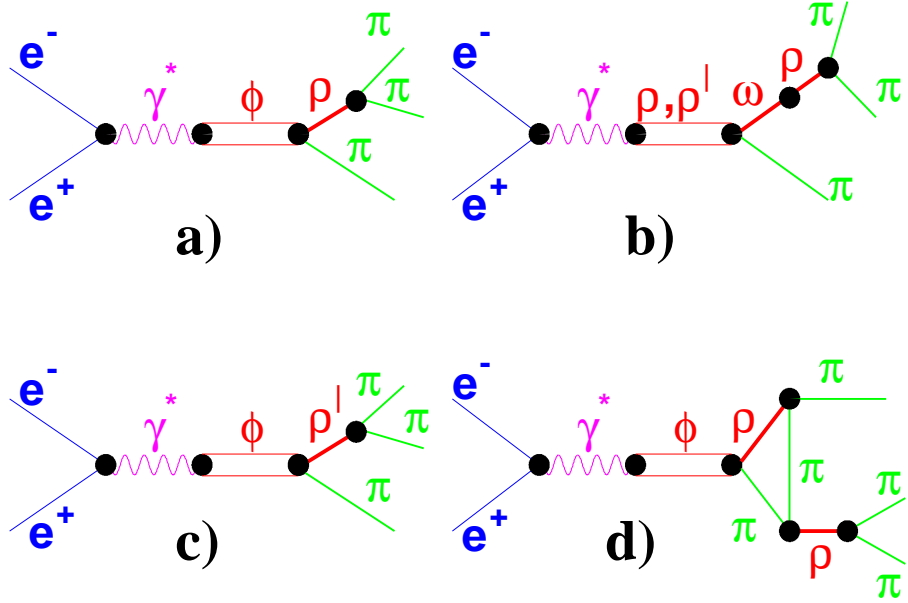,width=12cm}
\caption{$e^+e^-\to\phi\to\rho\pi\to\pi^+\pi^-\pi^0$ transition diagrams.
         a), b), c) -- through various intermediate states. d) -- the
	 $e^+e^-\to\phi\to\rho\pi\to\pi^+\pi^-\pi^0$ transition
	 with $\rho$ and $\pi$ mesons interaction in the final state.}
\label{dplfeim2}
\end{center}
\end{figure}

 The transition $V\to\pi^+\pi^-\pi^0$ besides $\rho\pi$ intermediate state
 (Fig.\ref{dplfeim2}a) can be performed by $\rho^{\prime(\prime\prime)}\pi$
 intermediate state (Fig.\ref{dplfeim2}c). Even if the branching ratio of the
 $V\to\rho^{\prime(\prime\prime)}\pi$ decay is rather small,
 the interference of this amplitude with the  $V\to\rho\pi$ amplitude 
 can give the noticeable contribution to the total cross section.
 The transition $e^+e^-\to\pi^+\pi^-\pi^0$ is also possible through 
 $\rho-\omega$ interference: $e^+e^-\to V\to\omega\pi^0\to\rho^0\pi^0$
 ($V=\rho,\rho^\prime,\rho^{\prime\prime}$) (Fig.\ref{dplfeim2}b). This effect
 was observed in the SND experiment in the energy region 
 $\sqrt[]{s}=1200$--$1400$ MeV \cite{sndro}.
 In the process $e^+e^-\to\rho\pi$ the interaction of the $\rho$-meson
 with pion in the final state is also expected \cite{akfaz}
 (Fig.\ref{dplfeim2}d).

 Besides of the $V \to 3\pi$ transition dynamics studies,
 the process  $e^+e^- \to \pi^+\pi^-\pi^0$ in the  $\phi(1020)$ resonance
 energy region can be used for the $\rho$-meson parameters measurement.
 In the reaction $e^+e^- \to \rho\pi$ the neutral and charged $\rho$-mesons
 are produced, so it can be used for the $\rho^\pm$ and $\rho^0$ mass
 difference determination \cite{razmass}. The $\rho$ mass values obtained by
 using different reactions contradict to each other \cite{pdg}. So it is 
 worthwhile to compare the resonance parameters ($m_\rho$ and $\Gamma_\rho$),
 obtained from the $e^+e^- \to \rho\pi$ reaction with the results of other
 experiments.

 Investigation of the $\pi\pi$ mass distribution in the process
 $e^+e^-\to\pi^+\pi^-\pi^0$ provides important information about vector
 mesons and their interference. Here we present the results of the dipion mass
 spectra analysis in SND experiment.

\section{Data analysis}

\subsection{Experiment}

 The SND detector \cite{sndnim} operated since 1995 up to 2000 at the VEPP-2M
 \cite{vepp2} collider in the energy range $\sqrt[]{s}$ from 360 to 1400 MeV.
 The detector contains several subsystems. The tracking system includes two
 cylindrical drift chambers. The three-layer spherical electromagnetic
 calorimeter is based on NaI(Tl) crystals. The muon/veto system consists of
 plastic scintillation counters and two layers of streamer tubes. The
 calorimeter energy and angular resolution depend on the photon energy as
 $\sigma_E/E(\%) = {4.2\% / \sqrt[4]{E(\mbox{GeV})}}$ and
 $\sigma_{\phi,\theta} = {0.82^\circ / \sqrt[]{E(\mathrm{GeV})}} \oplus 0.63^\circ$.
 The tracking system angular resolution is about $0.5^\circ$ and $2^\circ$ for
 azimuthal and polar angles respectively. The energy loss resolution $dE/dx$ in
 the drift chamber is about 30\% -- good enough to provide charged kaon
 identification in the $\phi$-meson production region.
 
 In 1998 SND had collected data in the energy region from 984 to 1060 MeV
 \cite{phi98} with integrated luminosity about $8.5~\mbox{pb}^{-1}$. For
 studies of the dipion mass spectra in the process
 $e^+e^- \to \pi^+\pi^-\pi^0$ the data sample collected in the energy region
 $\sqrt[]{s}=1016$ -- $1022$ MeV was used. The total integrated luminosity 
 accumulated in this region is about 4.3 pb$^{-1}$.

\subsection{Selection of $e^+e^- \to \pi^+\pi^-\pi^0$ events}

 The data analysis and selection criteria used in this work are similar
 to those described in Ref.\cite{phi98}. During the experimental runs, the
 first-level trigger \cite{sndnim} selects events with energy deposition
 in the calorimeter more then 200 MeV and  containing two or more charged
 particles. During processing of the experimental data the event
 reconstruction is performed \cite{sndnim,phi98}. For further analysis
 events containing two or more photons and two charged particles with
 $|z| < 10$ cm and $r < 1$ cm were selected. Here $z$ is the coordinate of the
 charged particle production point along the beam axis (the longitudinal size
 of the interaction region is $\sigma_z \sim 2.5$ cm); $r$ is the distance
 between the charged particle track and the beam axis in the $r-\phi$ plane.
 Extra photons in $e^+e^- \to \pi^+\pi^-\pi^0$ events can appear because of
 the beam background overlap or nuclear interactions of the charged pions in
 the calorimeter. Under these selection conditions, the background sources are
 $e^+e^- \to K^+K^-$, $K_SK_L$, $\eta\gamma$, $\omega\pi^0$,
 $e^+e^-\gamma\gamma$ processes and the beam background.

 To suppress the beam background, the following cuts on the angle between
 the two charged particles tracks $\psi$ and energy deposition of the neutral
 particles $E_{neu}$ were applied:
 $\psi > 40^{\circ}$, $E_{neu} > 0.1 \cdot \sqrt[]{s}$.

 To reject the background from the $e^+e^- \to K^+K^-$ process the following
 cuts were imposed: $(dE/dx) < 5 \cdot (dE/dx)_{min} $ for each charged
 particle, $(dE/dx) < 3 \cdot (dE/dx)_{min} $ at least for one of them, and
 $\Delta\phi < 10^\circ$. Here $|\Delta\phi|$ is an acollinearity angle in the
 azimuthal plane and $(dE/dx)_{min}$ is an average energy loss of a minimum
 ionizing particle. To suppress the  $e^+e^- \to e^+e^-\gamma\gamma$ events the
 energy deposition of the charged particles $E_{cha}$ was required to be
 small enough: $E_{cha} < 0.5 \cdot \sqrt[]{s}$.

\begin{figure}
\begin{center}
\epsfig{figure=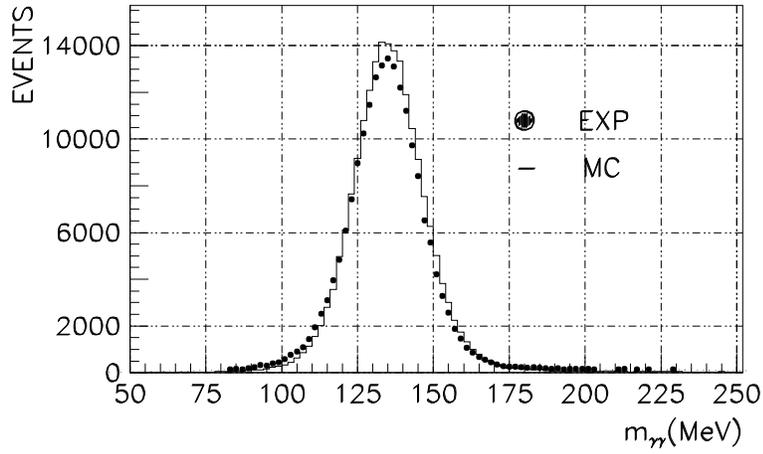,width=10cm}
\caption{Two-photon invariant mass distribution in the 
         $e^+e^- \to \pi^+\pi^-\pi^0$ events.}
\label{mgg}
\end{center}
\end{figure}

\begin{figure}
\begin{center}
\epsfig{figure=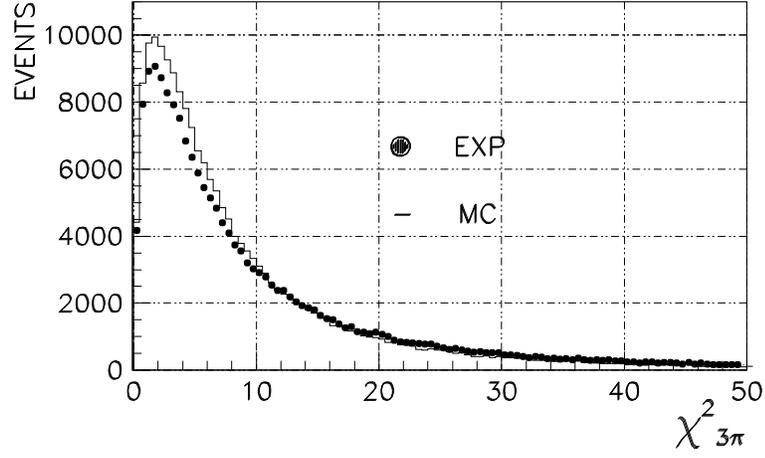,width=10cm}
\caption{The $\chi^2_{3\pi}$ distribution in $e^+e^- \to \pi^+\pi^-\pi^0$
         events.}
\label{chi}		  
\end{center}
\end{figure}

\begin{figure}
\begin{center}
\epsfig{figure=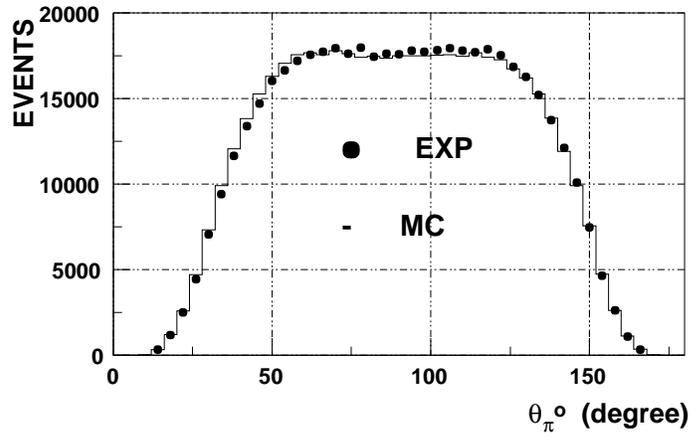,width=9cm}
\caption{The $\theta$ distribution of neutral pions from the reaction
         $e^+e^- \to \pi^+\pi^-\pi^0$.}
\label{dplteu3}
\end{center}
\end{figure}

\begin{figure}
\begin{center}
\epsfig{figure=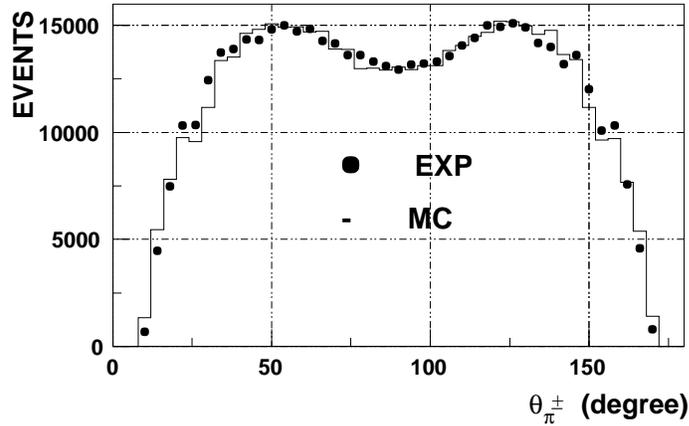,width=9cm}
\vspace{-1cm}
\caption{The $\theta$ distribution of charged pions from the reaction
         $e^+e^- \to \pi^+\pi^-\pi^0$.}
\label{dplteu12}
\end{center}
\end{figure}

\begin{figure}
\begin{center}
\epsfig{figure=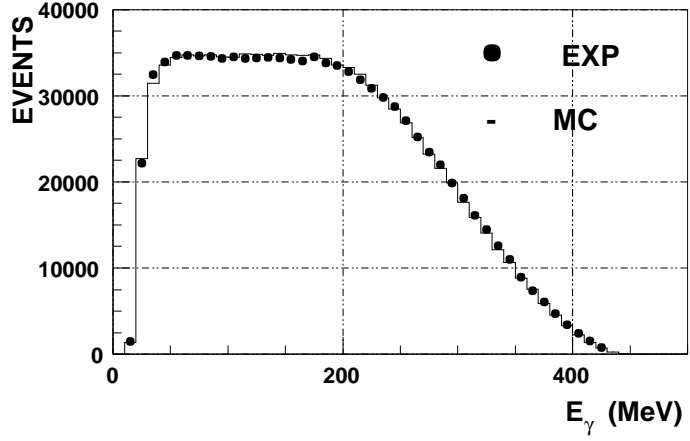,width=9cm}
\caption{Photons energy distribution.}
\label{dplepu45}
\end{center}
\end{figure}

\begin{figure}
\begin{center}
\epsfig{figure=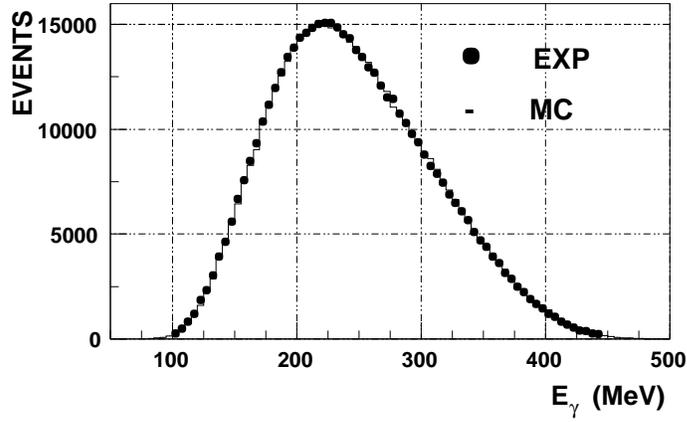,width=9cm}
\caption{The maximal photon energy distribution.}
\label{dplepu4}
\end{center}
\end{figure}

 The selected events were reconstructed under assumption that all registered
 particles have the common vertex in the $e^+e^-$ interaction region.
 The average position of the beams interaction point and its $rms$ spread were
 measured by using $e^+e^-\to e^+e^-$ events during data collection.
 Then a kinematic fit was
 performed under the following constraints: the charged particles are
 considered to be pions, the system has zero total momentum, the total energy
 is $\sqrt[]{s}$, and the photons originate from the $\pi^0 \to \gamma\gamma$
 decays. The value of the likelihood function $\chi^2_{3\pi}$ is calculated
 during the fit. In events with more than two photons, extra photons are
 considered as spurious ones and rejected. To do this, all possible subsets of
 two photons were inspected and the one, corresponding to the maximum
 likelihood was selected. The two-photon invariant mass and $\chi^2_{3\pi}$
 distributions are shown in Fig.\ref{mgg} and Fig.\ref{chi}. After the
 kinematic fit the following additional cuts were applied:
 $36^{\circ} < \theta_\gamma < 144^{\circ}$,  $N_{\gamma}=2$, and
 $\chi^2_{3\pi} < 20$. Here $\theta_\gamma$ is polar angle of one of the
 photons selected by the reconstruction program as originated from the
 $\pi^0$-decay, $N_{\gamma}$ is the number of detected photons. Under these
 criteria about $0.5 \times 10^6$ events were selected. In Fig.\ref{dplteu3}
 and Fig.\ref{dplteu12} the angular distributions of pions for the selected
 events are shown. While Fig.\ref{dplepu45} and Fig.\ref{dplepu4} demonstrates
 the photon energy distributions for the same events.

 The difference between described selection criteria and those used in
 Ref.\cite{phi98} is that the cuts on the pion and photon angles were
 relaxed here. This leads to the increase of the selected events number by a
 factor of about 1.5. The $\phi$ meson parameters obtained under this new
 selection criteria agree with results reported in the previous work.
 For example, the $e^+e^-\to\pi^+\pi^-\pi^0$ cross section changed by about
 0.5\%, what is negligible in comparison with the 5\% systematic error declared
 in Ref.\cite{phi98}. 

 The number of background events was estimated by using simulation in the
 following way:
\begin{eqnarray}
\label{bg}
  N_{bkg}({s}) = \sum_i \sigma_{Ri}({s}) \epsilon_i({s})  IL({s}),
\end{eqnarray}
 where $i$ is a process number, $\sigma_{Ri}({s})$ is the cross
 section of the background process taking into account the radiative
 corrections, $IL({s})$ is the integrated luminosity, $\epsilon_i({s})$ is
 the detection probability for the background process. The accuracy of
 the background events number determination is estimated to be about
 15 -- 20 \%. The numbers of $e^+e^- \to \pi^+\pi^-\pi^0$ events
 (after background subtraction) and background event numbers are shown in
 Table~\ref{tab1}. The $e^+e^- \to K_SK_L$ events are the main background
 source and their contribution is $\sim$ 1\%  of all selected events.

\subsection{Construction of the $\pi^+\pi^-$ and $\pi^\pm\pi^0$ mass spectra}

 For the selected events the $\pi^+\pi^-$ and $\pi^\pm\pi^0$ mass spectra
 were formed and arranged in histograms with dipion mass range from 280 to 880
 MeV and bin width of 20 MeV. All experimental energy points were used to
 produce a single invariant mass distribution. In case of the $\pi^\pm\pi^0$
 distribution the charged pion was selected at random from the two
 possibilities. The invariant mass values were calculated after the kinematic
 reconstruction.
 
 The expected background was subtracted bin by bin while forming the desired
 histograms. Simulated distributions were used for subtraction of the 
 $e^+e^- \to K^+K^-$, $\eta\gamma$, $\omega\pi^0$ and $e^+e^-\gamma\gamma$
 backgrounds. The $e^+e^- \to K_SK_L$ main background was studied using
 experimental events.

 As a rule, the background from the $e^+e^- \to K_SK_L$ originates when
 $K_S\to\pi^+\pi^-$ decay inside the collider vacuum chamber is accompanied by
 ``photons'' in calorimeter produced due to $K_L$ meson nuclear interactions
 or its decay in flight inside the calorimeter. To select $K_SK_L$ events for
 related background studies the following additional cuts were applied
 $r > 0.2$ cm for both charged particles and $\psi>130^\circ$. Note that $K_S$
 decay length is about 0.5 cm for our energies and the angle $\psi$ between
 the pions from the $K_S$ decay is about $150^\circ$.

\begin{figure}
\begin{center}
\epsfig{figure=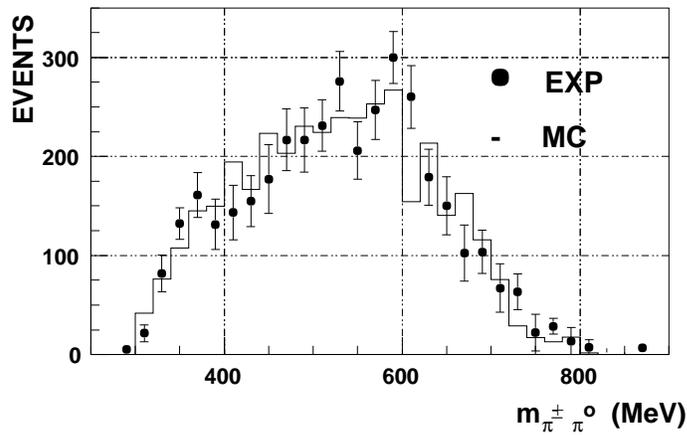,width=9cm}
\caption{``$\pi^\pm\pi^0$'' mass distribution from $e^+e^- \to K_SK_L$ 
         background events.} 
\label{dplkslcha}	 
\end{center}
\end{figure}

\begin{figure}
\begin{center}
\epsfig{figure=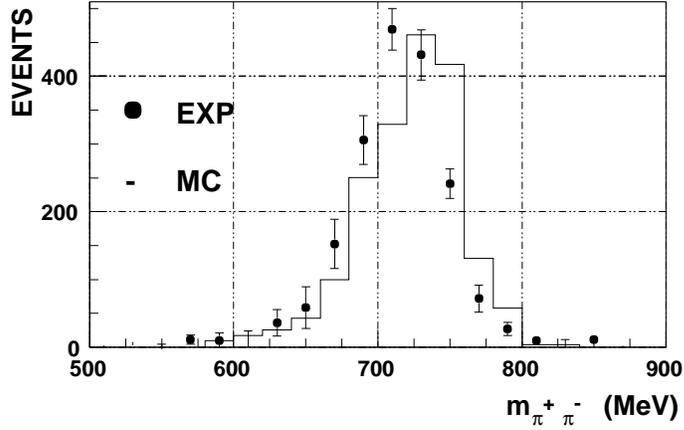,width=9cm}
\caption{ ``$\pi^+\pi^-$'' mass distribution from the $e^+e^- \to K_SK_L$ 
         background events.}
\label{dplkslneu}
\end{center}
\end{figure}

\begin{figure}
\begin{center}
\epsfig{figure=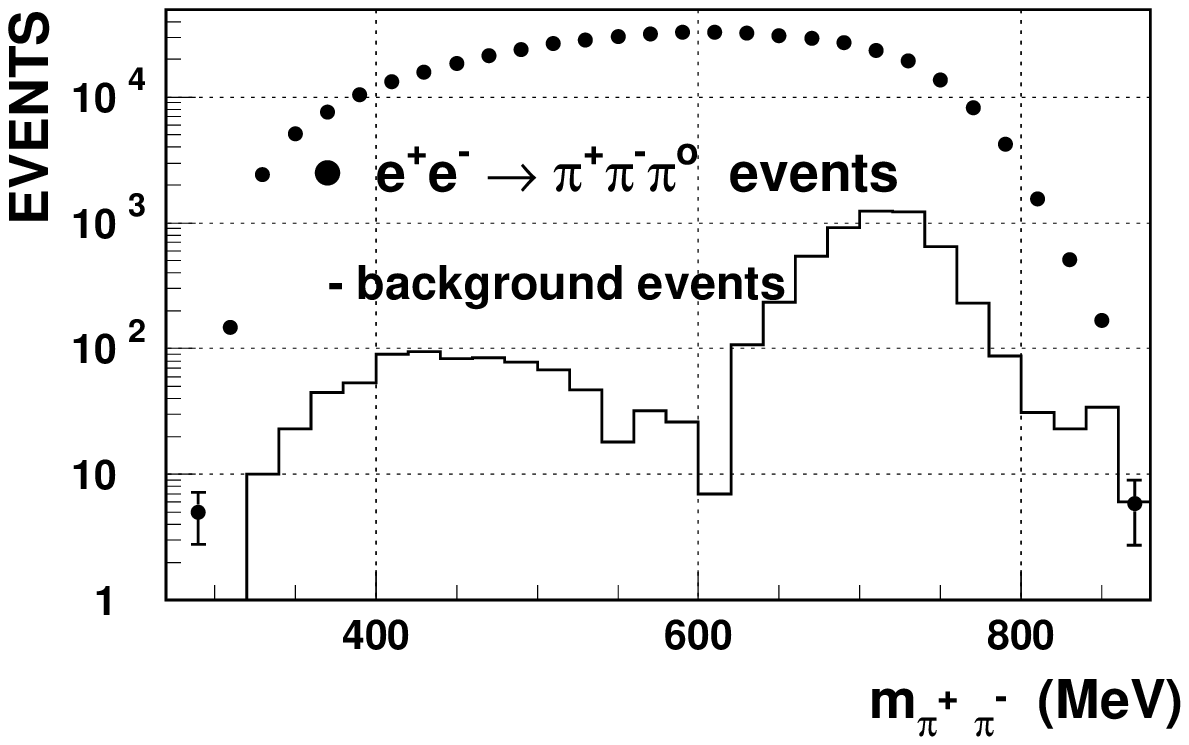,width=10cm}
\caption{$\pi^+\pi^-$ mass distribution for selected
         $e^+e^- \to \pi^+\pi^-\pi^0$ events and background contribution}
\label{dpleffon0}
\end{center}
\end{figure}

\begin{figure}
\begin{center}
\epsfig{figure=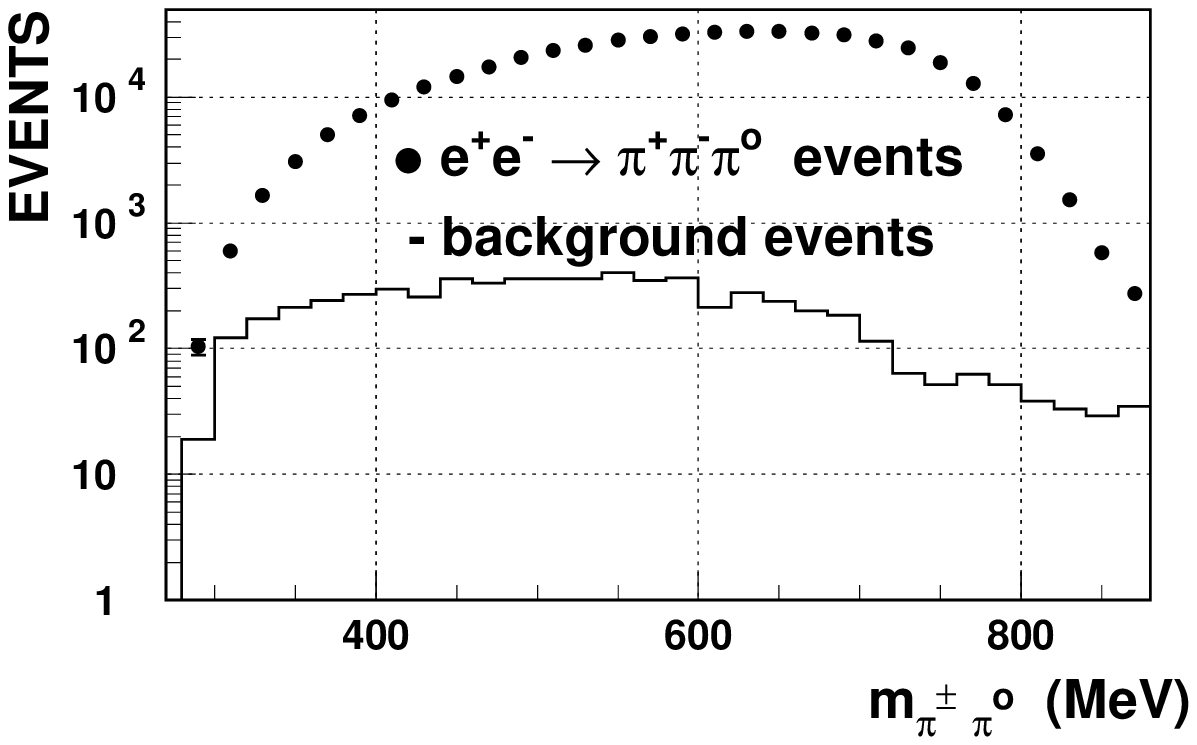,width=10cm}
\caption{$\pi^\pm\pi^0$ mass distribution for selected
         $e^+e^- \to \pi^+\pi^-\pi^0$ events and background contribution}
\label{dpleffonc}
\end{center}
\end{figure}

 The experimental and simulated ``$\pi^\pm\pi^0$'' distributions are consistent
 to each other (Fig.\ref{dplkslcha}), so the simulated distribution was used
 for the $K_SK_L$ background subtraction in the $\pi^\pm\pi^0$ invariant mass
 spectrum. The agreement is not so good for the ``$\pi^+\pi^-$'' mass spectrum:
 the simulated histogram is shifted from the experimental one by about 3\% in
 the peak region (Fig.\ref{dplkslneu}). This disagreement is caused due to
 inaccuracies in simulation of the $K_L$ nuclear interactions. The experimental
 distribution (Fig.\ref{dplkslneu}) was used for $K_SK_L$ background
 subtraction for events in which both charged particles had $r>0.2$ cm.
 Unfortunately, this procedure is not justified if at least one charged
 particle in the event has $r<0.2$ cm. Indeed, ``$\pi^+\pi^-$'' mass
 distribution for the $K_SK_L$ background depends crucially on the shape of
 the angle $\psi$ distribution, which in its turn depends on the value of $r$,
 because during event reconstruction it was assumed that all charged particle
 tracks have a common vertex in the beams interaction region. For the
 $K_S\to\pi^+\pi^-$ vertex this assumption is not true. Therefore,
 ``$\pi^+\pi^-$'' distribution obtained for $K_SK_L$ background under
 condition $r>0.2$ cm cannot be used for subtraction of all $K_SK_L$
 background. Instead for events with $r<0.2$ cm for one of the charged
 particles the simulated distribution, corrected for the above mentioned 3\%
 discrepancy, was used for $K_SK_L$ background subtraction. The dipion invariant
 mass spectra and background contributions obtained in such a way are shown in 
 Fig.\ref{dpleffon0}, \ref{dpleffonc}.

\section{Theoretical framework}

 In the VDM framework the cross section of the  $e^+e^-\to\pi^+\pi^-\pi^0$
 process is
\begin{eqnarray}
\label{ds}
 {d\sigma \over dm_0 dm_+} = { {4\pi\alpha} \over {s^{3/2}} }
 \biggl|A_{\rho\pi}(s)\biggr|^2
 {{|\vec{p}_+ \times \vec{p}_-|^2} \over {12\pi^2\mbox{~}\sqrt[]{s}}} m_0m_+ \cdot
 |F|^2,
\end{eqnarray}
 where $\vec{p}_+$ and $\vec{p}_-$ are the $\pi^+$ and $\pi^-$
 momenta, $m_0$ and $m_+$ are $\pi^+\pi^-$ and $\pi^+\pi^0$ invariant masses. 
 Formfactor $F$ of the $\gamma^\star \to \pi^+\pi^-\pi^0$ transition
 has the form
\begin{eqnarray}
\label{formfac}
 |F|^2 = \Biggl| { g_{\rho^0\pi\pi} \over D_\rho(m_0) Z(m_0)} +
 {g_{\rho^+\pi\pi} \over D_{\rho^+}(m_+) Z(m_+)} +
 {g_{\rho^-\pi\pi} \over D_{\rho^-}(m_-) Z(m_-)} + 
 {A_{\omega\pi}(s) \over A_{\rho\pi}(s)}
 {\Pi_{\rho\omega}g_{\rho^0\pi\pi}\over D_\rho(m_0) D_\omega(m_0)} +
 a_{3\pi} \Biggr|^2
\end{eqnarray}
 Here $$D_\rho(m_0) = m_{\rho^0}^2 - m_0^2 -im_0\Gamma_{\rho^0}(m_0),\mbox{~~}
 D_{\rho^\pm}(m_\pm) = m_{\rho^\pm}^2 - m_\pm^2 -im_\pm\Gamma_{\rho^\pm}(m_\pm)$$
 $$\Gamma_{\rho^0}(m_0) = \Biggl({m_{\rho^0} \over m_0}\Biggr)^2
   \Gamma_{\rho^0} \Biggl({q_0(m_0) \over q_0(m_{\rho^0})}\Biggr)^3, \mbox{~~}
   \Gamma_{\rho^\pm}(m_\pm) = \Biggl({m_{\rho^\pm} \over m_\pm}\Biggr)^2
   \Gamma_{\rho^\pm} \Biggl({q_\pm(m_\pm) \over q_\pm(m_{\rho^\pm})}\Biggr)^3,$$
 $$q_0(m) = {1 \over 2}(m^2-4m_\pi^2)^{1/2},
 q_\pm(m) = {1 \over 2m}
 ((m^2-(m_{\pi^0}+m_\pi)^2)(m^2-(m_{\pi^0}-m_\pi)^2)^{1/2},$$
 $$m_-=\sqrt[]{s+m_{\pi^0}^2+2m_{\pi}^2-m_0^2-m_+^2},$$
 where $m_-$ is $\pi^-\pi^0$ invariant mass, $m_{\pi^0}$ and $m_\pi$ are the
 neutral and charged pion masses.
 The $\rho^0 \to \pi^+\pi^-$ and $\rho^\pm \to \pi^\pm\pi^0$ transition
 coupling constants could be determined in the following way:
 $$g_{\rho^0\pi\pi}^2 = {6\pi m_{\rho^0}^2\Gamma_{\rho^0} \over
 q_0(m_{\rho^0})^3}, \mbox{~~}
 g_{\rho^\pm\pi\pi}^2 = {6\pi m_{\rho^\pm}^2\Gamma_{\rho^\pm} \over
 q_\pm(m_{\rho^\pm})^3}$$
 If these constants are equal, then the $\rho^0$ and $\rho^\pm$ meson
 widths are related as follows:
\begin{eqnarray}
\label{shir}
 \Gamma_{\rho^\pm} = \Gamma_{\rho^0}{m_{\rho^0}^2 \over m_{\rho^\pm}^2}
 { q_\pm(m_{\rho^\pm})^3 \over q_0(m_{\rho^0})^3}
\end{eqnarray}

\begin{figure}
\begin{center}
\epsfig{figure=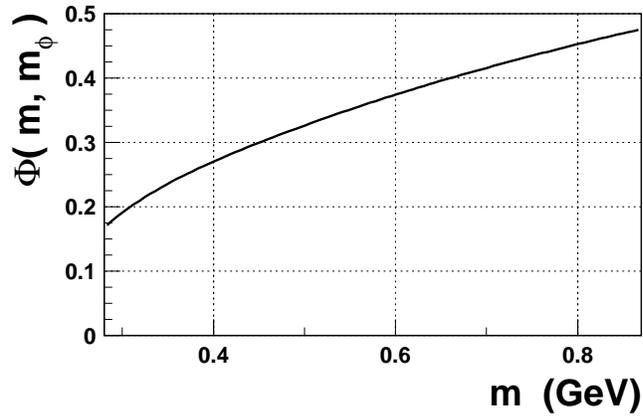,width=10cm}
\caption{$\Phi(m,s)$ as the function of the dipion mass $m$ at
         $\sqrt[]{s} = m_\phi$}
\label{dplfaza}
\end{center}
\end{figure}

 Factor $Z(m) = 1 -is_1\Phi(m,s)$ takes into account the  interaction of the
 $\rho$ and $\pi$ mesons in the final state \cite{akfaz}. $\Phi(m,s)$ as a
 function of the pion pair invariant mass $m$ is shown in Fig.\ref{dplfaza}
 for the $e^+e^-$ center-of-mass energy $\sqrt[]{s}=m_\phi$. The free parameter
 in our fit $s_1$ was introduced to account for possible deviations from the
 Ref.\cite{akfaz} prediction $s_1=1$. The fourth item in (\ref{formfac}) takes
 into account the $\rho-\omega$ mixing \cite{rhoom}. Polarization operator of
 these mixing $\Pi_{\rho\omega}$ satisfies  
 $\mbox{Im}(\Pi_{\rho\omega}) \ll \mbox{Re}(\Pi_{\rho\omega})$, where
 $\mbox{Re}(\Pi_{\rho\omega}) = 2m_\omega\delta$ with $\delta = 2.3$ MeV
 \cite{akfaz,akozi}. So we have assumed $\mbox{Im}(\Pi_{\rho\omega}) = 0$ in
 subsequent analysis. Amplitudes of the $\gamma^\star \to \rho\pi$ and
 $\gamma^\star \to \omega\pi^0$ transitions have the form
\begin{eqnarray}
 A_{\rho\pi}(s) = \sum_{V=\omega,\phi,\omega^\prime,\omega^{\prime\prime}}
 {g_{\gamma V}g_{V\rho\pi} \over D_V(s)}, \mbox{~~}
 A_{\omega\pi}(s) = \sum_{V=\rho,\rho^\prime,\rho^{\prime\prime}}
 {g_{\gamma V}g_{V\omega\pi^0} \over D_V(s)},
\end{eqnarray}
 Using SND measurements of the $e^+e^-\to\pi^+\pi^-\pi^0$ and
 $e^+e^- \to \pi^0\pi^0\gamma$ cross sections \cite{phi98,ppg}, we can
 express the ratio of these amplitudes in the $\phi$-meson region as
 $$ A_{\omega\pi}(s)/A_{\rho\pi}(s) =
 K_{\omega\pi/\rho\pi}e^{i\phi_{\omega\pi/\rho\pi}},
 K_{\omega\pi/\rho\pi} \simeq 0.3,
 \phi_{\omega\pi/\rho\pi} \simeq -110^\circ,
 \sqrt[]{s}=m_\phi.
 $$
 The amplitude $a_{3\pi}$ takes into account possible additional intermediate
 states in the $\gamma^\star \to \pi^+\pi^-\pi^0$ transition. For example,
 $\gamma^\star \to \rho^\prime(\rho^{\prime\prime})\pi \to \pi^+\pi^-\pi^0$
 transition will lead to the contribution
\begin{eqnarray}
\label{a3pi}
 a_{3\pi} = {{A_{\rho^\prime\pi}(s)} \over {A_{\rho\pi}}(s)}
 \Biggl( {g_{\rho^\prime\pi\pi} \over D_{\rho^\prime}(m_0)} +
 {g_{\rho^\prime\pi\pi} \over D_{\rho^\prime}(m_+)} +
 {g_{\rho^\prime\pi\pi} \over D_{\rho^\prime}(m_-)} \Biggr),
\end{eqnarray}
 where
\begin{eqnarray}
 A_{\rho^\prime\pi}(s) = \sum_{V=\omega,\phi,\omega^\prime}
 {g_{\gamma V}g_{V\rho^\prime\pi} \over D_V(s)}.
\end{eqnarray}
 The ratio $A_{\rho^\prime\pi} / A_{\rho\pi}$ is expected to be real for
 energies $\sqrt[]{s} \simeq m_\phi$ \cite{copco} where
 $${A_{\rho^\prime\pi} \over A_{\rho\pi}} \simeq
 {g_{\phi\rho^\prime\pi}\over g_{\phi\rho\pi}}. $$
 So the imaginary part of  (\ref{a3pi}) is negligible due to the large
 $\rho^\prime$ mass ($m_{\rho^\prime} \sim 1400$ MeV). Therefore the $a_{3\pi}$
 amplitude was assumed to be real in our analysis.

\section{Approximation of the $\pi\pi$ mass spectra}

\subsection{Theoretical distributions}

\begin{figure}
\begin{center}
\epsfig{figure=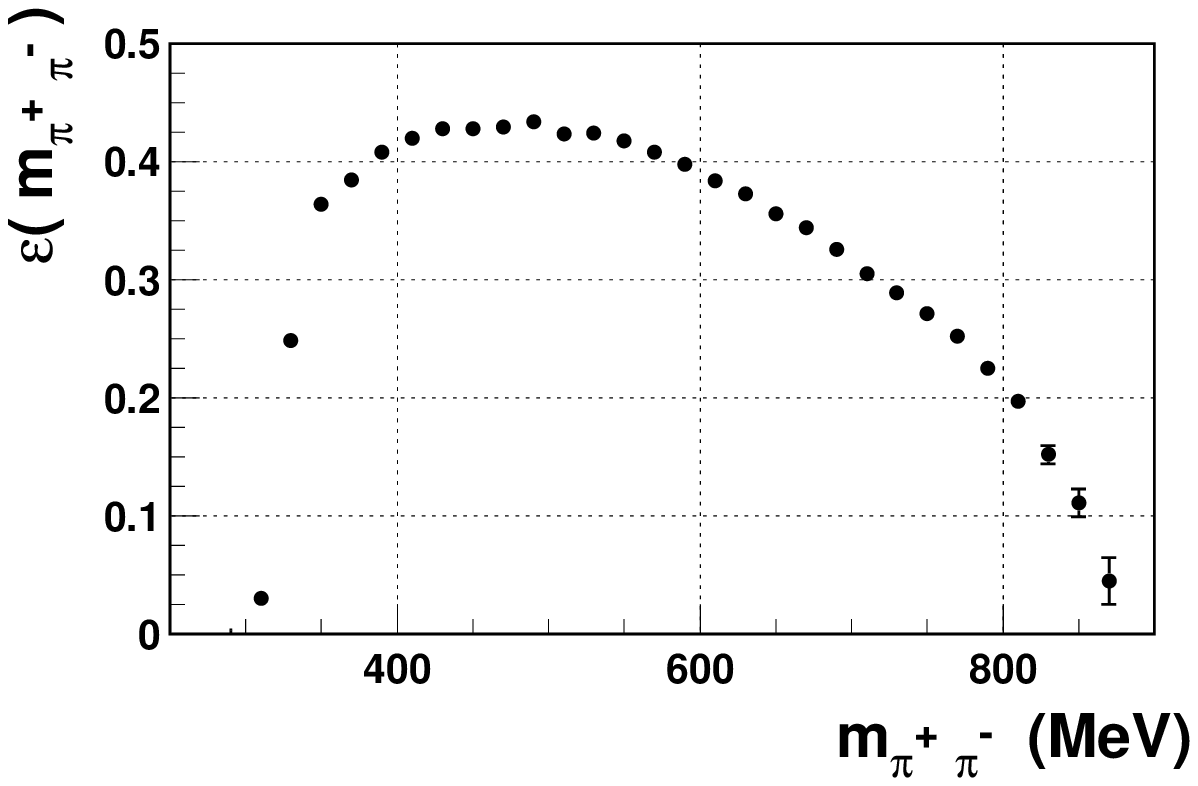,width=10cm}
\caption{Detection efficiency for $\pi^+\pi^-$ mass spectrum.}
\label{dpleffm0}
\end{center}
\end{figure}

\begin{figure}
\begin{center}
\epsfig{figure=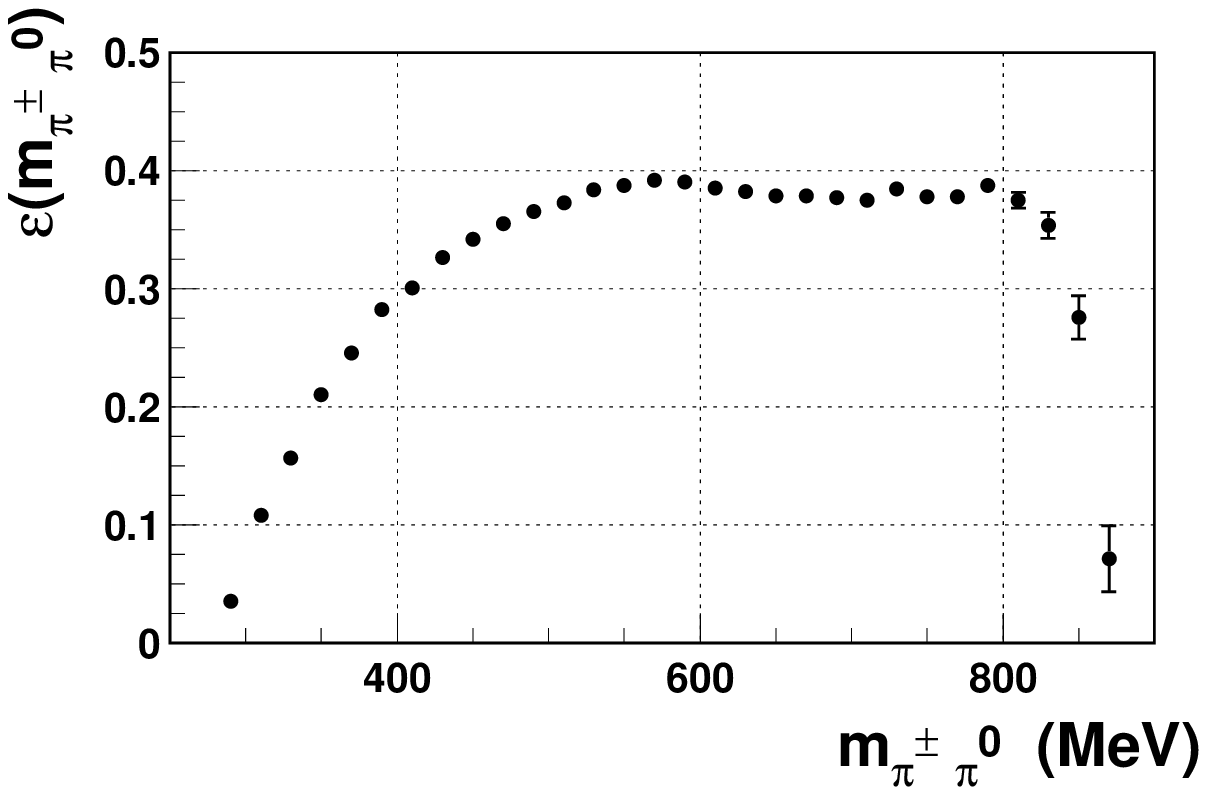,width=10cm}
\caption{Detection efficiency for $\pi^\pm\pi^0$ mass spectrum.}
\label{dpleffmc}
\end{center}
\end{figure}

 The experimental dipion mass spectra (Tables \ref{tab7} and \ref{tab8})
 were fitted with theoretical distributions. Using the 
 $e^+e^-\to\pi^+\pi^-\pi^0$ cross section (\ref{ds}) the
 theoretical spectra were calculated in the following way:
\begin{eqnarray}
 S^{(0)}_j(s) = {1 \over C_S(s)} \cdot \int\limits^{m_{j}+\Delta}_{m_{j}-\Delta}
 m_0 dm_0 \int\limits^{m_+^{max}(m_0)}_{m_+^{min}(m_0)}
 m_+ dm_+ |\vec{p}_+ \times \vec{p}_-|^2 \cdot |F|^2,
\end{eqnarray}
\begin{eqnarray}
 S^{(\pm)}_j(s) = {1 \over C_S(s)} \cdot \int\limits^{m_{j}+\Delta}_{m_{j}-\Delta}
 m_\pm dm_\pm \int\limits^{m_0^{max}(m_+)}_{m_0^{min}(m_+)}
 m_0 dm_0 |\vec{p}_+ \times \vec{p}_-|^2 \cdot |F|^2,
\end{eqnarray}
 where $j$ is the bin number, $\Delta=10$ MeV - half of the bin width,
 $m_j$ - the central value of the invariant mass in the $j$th bin, $C_S(s)$
 - normalizing coefficient. These spectra were corrected taking into account
 the detection efficiency $\epsilon_j$ (Fig.\ref{dpleffm0}~and~\ref{dpleffmc})
 for the $j$th bin and a probability $a_{ij}$ for the event belonging to the 
 $j$th bin to migrate to the $i$th bin due to finite detector resolution
\begin{eqnarray}
 G^{(0)}_i(s) = \Biggl(\sum_j a^{(0)}_{ij} S^{(0)}_j(s)
 \epsilon^{(0)}_j \Biggr) \cdot (1+\delta^{(0)}_i(s)) \cdot
 \beta^{(0)}_i
\end{eqnarray}
\begin{eqnarray}
 G^{(\pm)}_i(s) = \Biggl(\sum_j a^{(\pm)}_{ij} S^{(\pm)}_j(s)
 \epsilon^{(\pm)}_j \Biggr) \cdot (1+\delta^{(\pm)}_i(s)) \cdot
 \beta^{(\pm)}_i
\end{eqnarray}
 Here $\delta_i(s)$ is a radiative correction, $\beta^{(0)}_i$ and
 $\beta^{(\pm)}_i$ are corrections due to inaccuracies in simulation of the 
 nuclear interactions of the charged pions. The values of $a_{ij}$,
 $\epsilon_{j}$ and $\delta_i(s)$ were obtained from simulation.
 
 These distributions were formed for all energy points $\sqrt[]{s}$
 (Table \ref{tab1}) and then summed as follows:
\begin{eqnarray}
\label{tg}
  P_i^{(0)} = \sum_{s} w(s) \cdot { G_i^{(0)}(s) \over C_G(s)},\mbox{~~}
  P_i^{(\pm)} = \sum_{s} w(s) \cdot { G_i^{(\pm)}(s) \over C_G(s)}
\end{eqnarray}
 Here $C_G(s)=\sum_{i}G_i(s)$ is a normalization factor, $w(s)= {N(s)/ N}$ is 
 the weight factor, $N(s)$ and $N$ being the numbers of the
 $e^+e^-\to\pi^+\pi^-\pi^0$ events in each energy point and the total events
 number respectively (Table \ref{tab1}).

\subsection{Detection efficiency}

 To obtain the $a_{ij}$, $\epsilon_{j}$ and $\delta_i(s)$, the Monte-Carlo
 sample of about $1.4 \times 10^6$ $e^+e^-\to\rho\pi\to\pi^+\pi^-\pi^0$
 events were processed in the same way as experimental data. The statistical
 errors of $a_{ij}$ coefficients were included in the error of $P_i$
 (Tables.\ref{tab7} and \ref{tab8}). The average detection efficiency obtained
 by simulation is about 0.37 and varies from 0.05 to 0.4 depending on
 dipion mass value (Fig.~\ref{dpleffm0}~and~\ref{dpleffmc}).

\begin{figure}
\begin{center}
\epsfig{figure=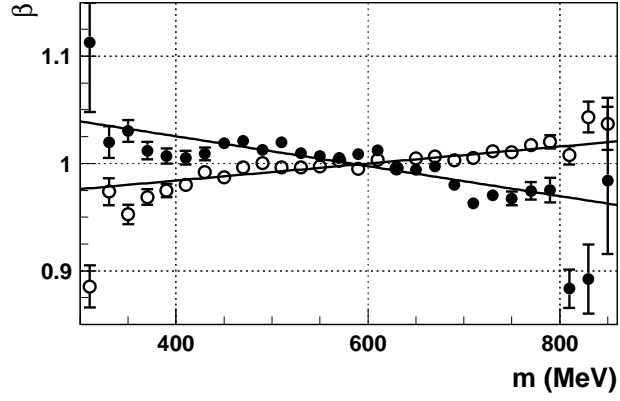,width=10cm}
\caption{The correction coefficients $\beta^{(0)}$ and $\beta^{(\pm)}$ upto
         irrelevant normalization factor. Dots -  $\beta^{(0)}$ correction
	 coefficient to the detection efficiency for $\pi^+\pi^-$ mass
	 spectrum. Circles -- $\beta^{(\pm)}$ correction coefficient to the
	 detection efficiency for $\pi^\pm\pi^0$ mass spectrum}
\label{dplpop}
\end{center}
\end{figure}

 Inaccuracies in the $\chi^2_{3\pi}$ and $dE/dx$ simulations lead to an error
 in average detection efficiency \cite{phi98}, but do not modify its dependence
 on the $\pi\pi$ invariant mass, except the regions of $m_{\pi\pi}<315$ MeV and
 $m_{\pi\pi}>845$ MeV. The cut $N_\gamma=2$ is also a source of some error in
 the detection efficiency determination, due to inaccuracies in the extra
 photons simulation. As was mentioned above, extra photons can appear because
 of the beam background overlap or charged pion nuclear interactions. In fact,
 if the extra photons originate from the beam background overlap, $N_\gamma=2$
 criteria  does not modify the detection efficiency dependence on a dipion
 mass, because the probability of beam background overlap does not depend on
 dipion mass. On the contrary, the probability of extra photons appearance
 because of nuclear interactions of the charged pions does depend on pion
 energy. Therefore any inaccuracy in simulation of these nuclear interactions
 will transform in inaccuracies in the simulation of extra photons and can
 lead to the detection efficiency error which will depend on the dipion mass.
 To correct this error we used the correction factor obtained in the following
 way:
\begin{eqnarray}
 \beta_i^{(0)} = {{n^{(0)}_i/N^{(0)}_i} \over {k_i^{(0)}/K_i^{(0)}}}, \mbox{~~}
 \beta_i^{(\pm)} = {{n^{(\pm)}_i/N^{(\pm)}_i} \over {k^{(\pm)}_i/K^{(\pm)}_i}},
\end{eqnarray}
 where $K_i$ and $N_i$ are the event numbers in the $i$th bin of experimental
 and simulated mass spectra respectively under condition $N_\gamma \ge 2$.
 $k_i$ and $n_i$ are the event numbers when the cut $N_\gamma = 2$ was imposed.
 The dependence of these correction factors on the dipion mass
 (Fig.\ref{dplpop}) can be approximated by linear functions:
$$
 \beta^{(0)}_i = C^{(0)} \cdot (1+a_0 \cdot m_i), \mbox{~~}
 \beta^{(\pm)}_i = C^{(\pm)} \cdot (1+a_\pm \cdot m_i)
$$
 with slopes $a_0=-0.13\cdot 10^{-3}$ and $a_\pm=0.083 \cdot 10^{-3}$ for
 $\pi^+\pi^-$ and $\pi^\pm\pi^0$ pairs respectively. As was mentioned above,
 the detection efficiency in the regions $m_{\pi\pi}<315$ MeV and
 $m_{\pi\pi}>845$ MeV have a large error. Uncertenity of $P_i$ in these
 regions was increased by addition of the 100\% spread.
 
\subsection{Fitting of the experimental distributions}

 The $\pi^+\pi^-$ and $\pi^\pm\pi^0$ mass spectra were fitted together. The
 function to be minimized was $\chi^2=\chi^2_0+\chi^2_\pm$, where
\begin{eqnarray}
 \chi^2_{0} = \sum_{i} \Biggl(
 { {H_i^{(0)}-P_i^{(0)}} \over {\sigma_i^{(0)}} }\Biggr)^2, \mbox{~~}
 \chi^2_{\pm} = \sum_{i} \Biggl(
 { {H_i^{(\pm)}-P_i^{(\pm)}} \over {\sigma_i^{(\pm)}} }\Biggr)^2 .
\end{eqnarray}
 Here $H^{(0)}$ and $H^{(\pm)}$ are the normalized  $\pi^+\pi^-$ and
 $\pi^\pm\pi^0$ mass distributions (histograms);
 $\sigma_i^{(0)}=\Delta H^{(0)}_i \oplus \Delta P^{(0)}_i$ and
 $\sigma_i^{(\pm)}=\Delta H^{(\pm)}_i \oplus \Delta P^{(\pm)}_i$ include 
 the uncertainties $\Delta H_i$ and $\Delta P_i$ of the experimental and
 theoretical distributions.

\begin{figure}
\begin{center}
\epsfig{figure=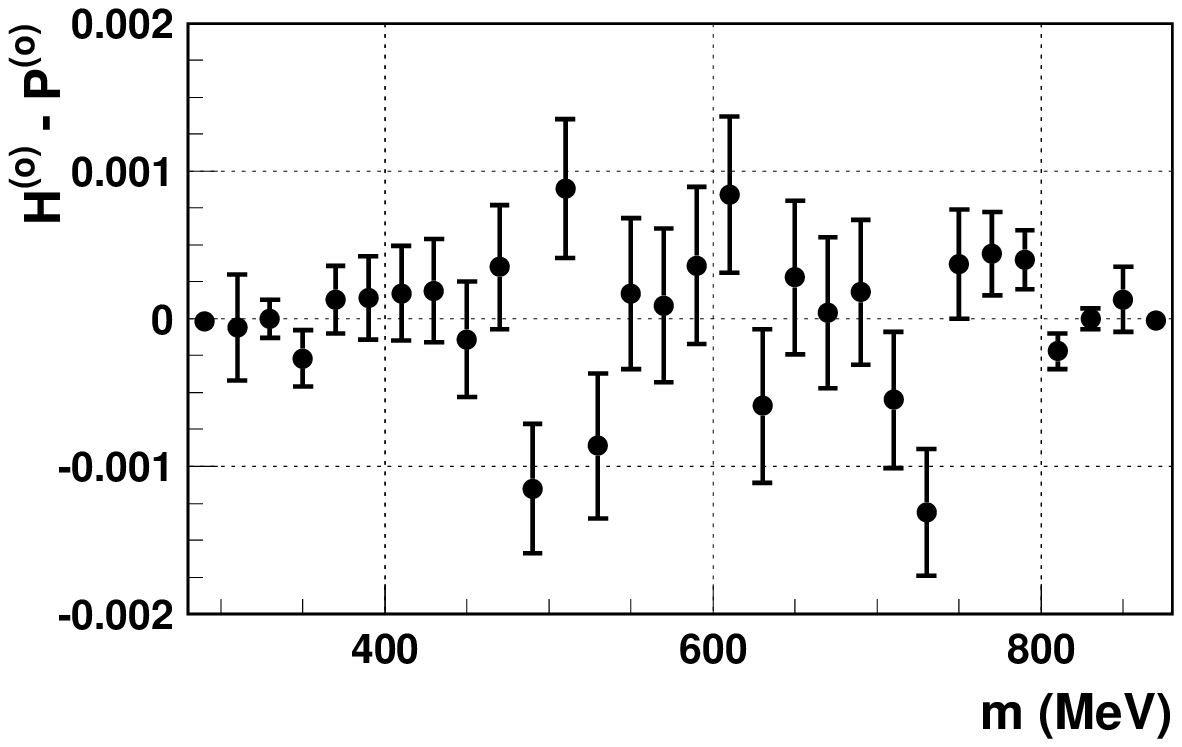,width=10cm}
\caption{The difference between the experimental and theoretical
         $m_{\pi^+\pi^-}$ spectra.}
\label{dplexte0d}
\end{center}
\end{figure}

\begin{figure}
\begin{center}
\epsfig{figure=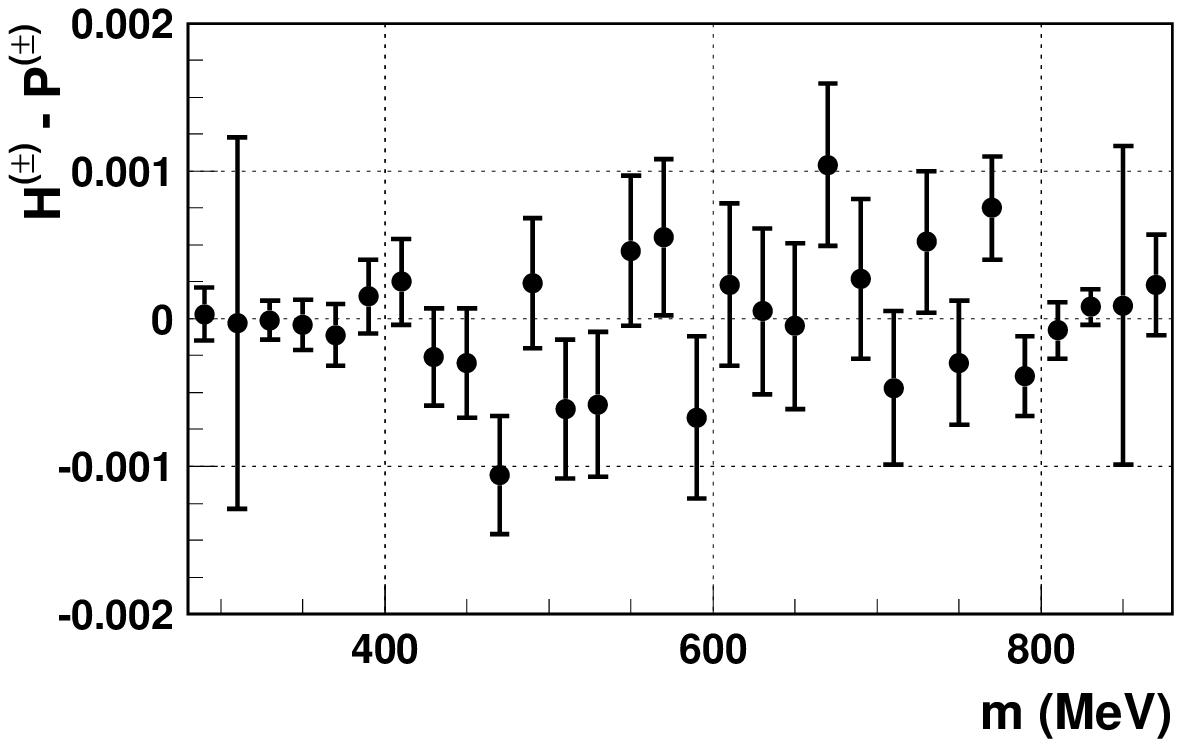,width=10cm}
\caption{The difference between the experimental and theoretical
         $m_{\pi^\pm\pi^0}$ spectra.}
\label{dplextecd}
\end{center}
\end{figure}

\begin{figure}
\begin{center}
\epsfig{figure=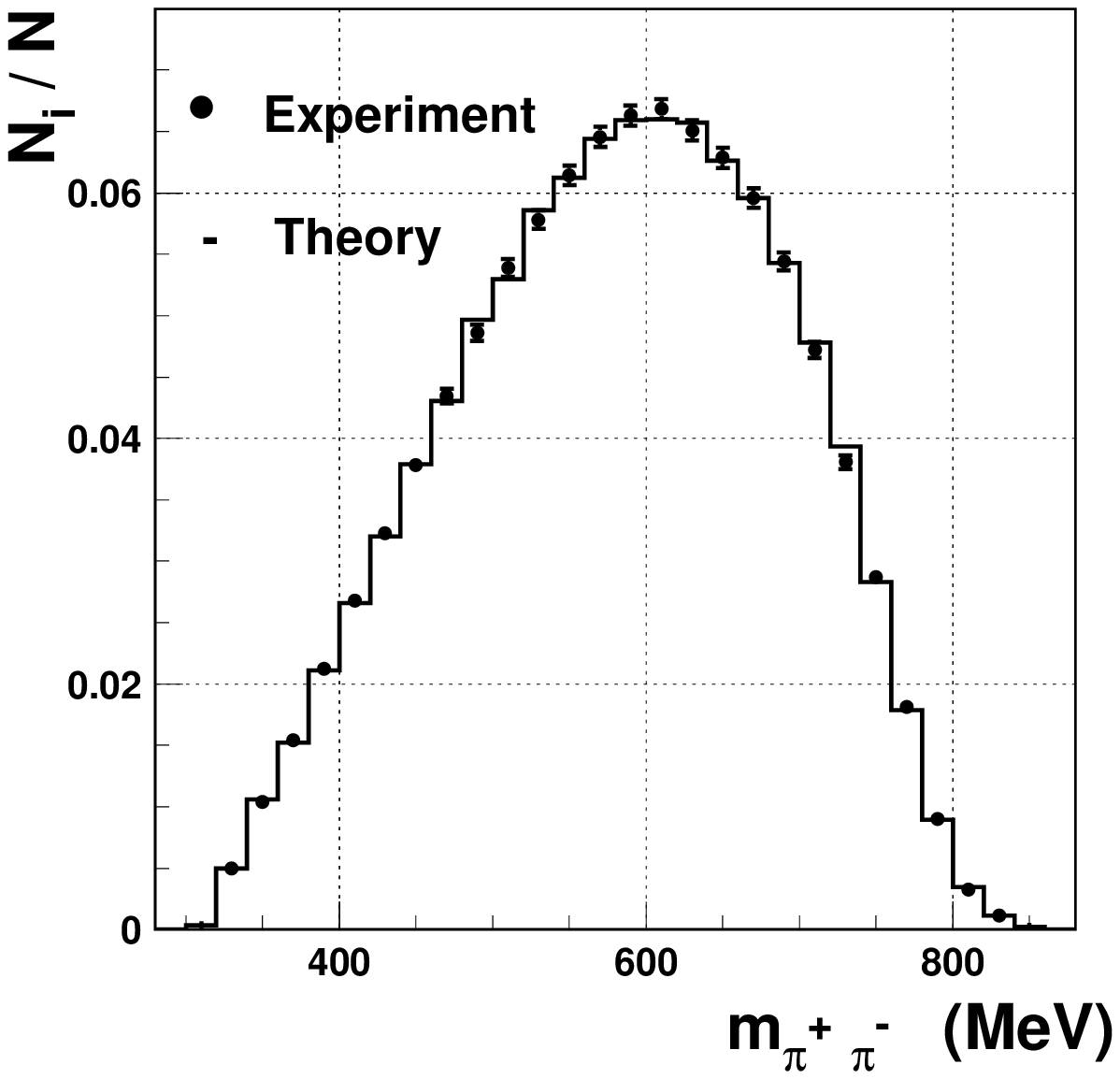,width=8.5cm}
\caption{The fit result for the $\pi^+\pi^-$ mass spectrum.}
\label{dplspe1}
\end{center}
\end{figure}

\begin{figure}
\begin{center}
\epsfig{figure=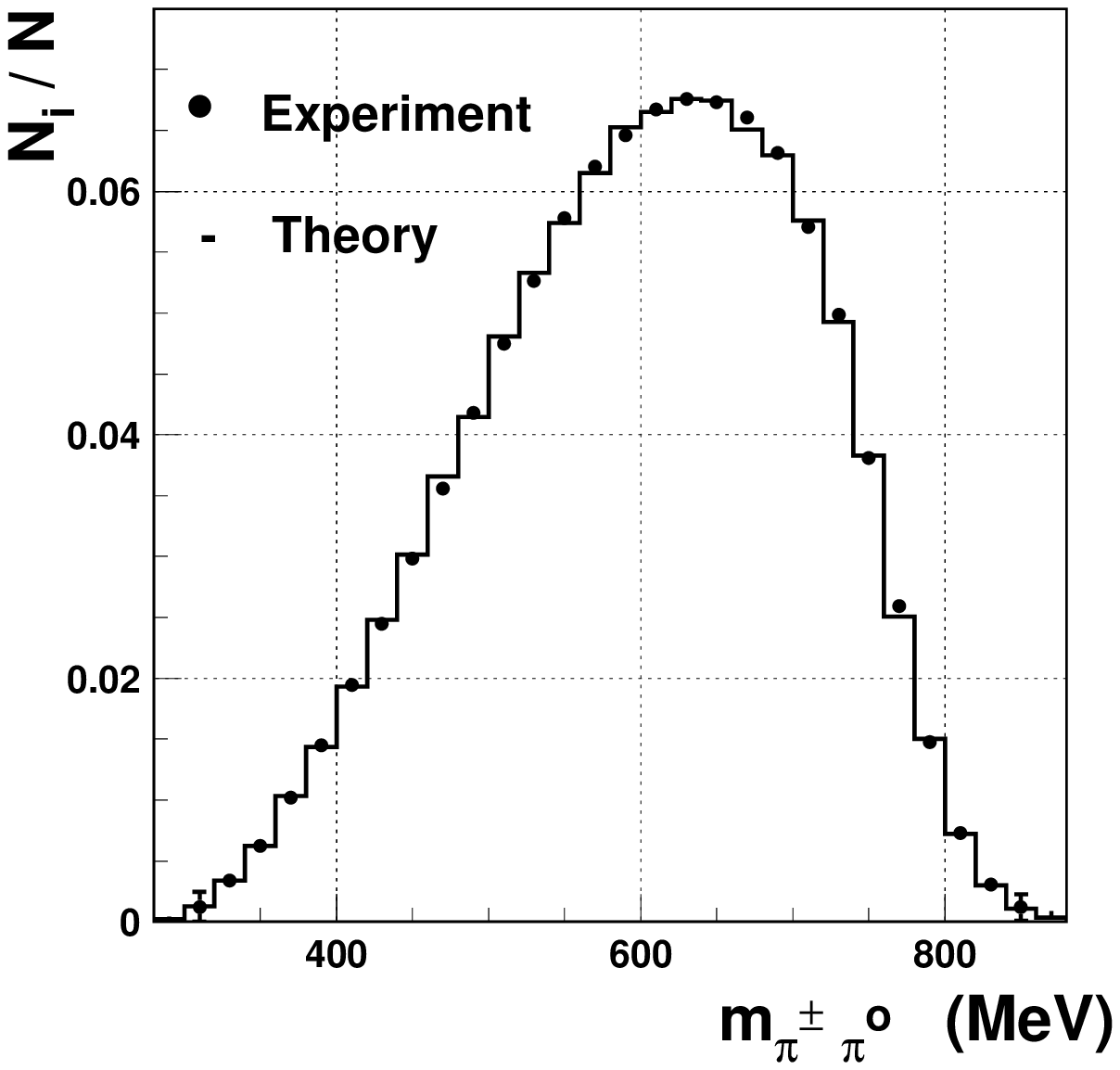,width=8.5cm}
\caption{The fit result for the $\pi^\pm\pi^0$ mass spectrum.}
\label{dplspe2}
\end{center}
\end{figure}

 The fitting was performed with $m_{\rho^0}$, $m_{\rho^\pm}-m_{\rho^0}$,
 $\Gamma_{\rho^0}$, $\Gamma_{\rho^\pm}$ as free parameters under the
 following assumptions:
\begin{enumerate}
\item
 $a_{3\pi} = 0$, $K_{\omega\pi/\rho\pi} = 0$, $s_1=0$;
\item
 $a_{3\pi}$ was free parameter,
 $K_{\omega\pi/\rho\pi} = 0$, $s_1=0$;
\item
 $a_{3\pi} = 0$,
 $K_{\omega\pi/\rho\pi}$ and $\phi_{\omega\pi/\rho\pi}$ were free parameters,
 $s_1=0$;
\item
 $a_{3\pi} = 0$, $K_{\omega\pi/\rho\pi} = 0$,
 $s_1$ was free parameter.
\end{enumerate}
 The $\chi^2$ value for all variants approximately equals to  $73$ for 54 -- 52
 degrees of freedom, besides $\chi^2_0 \simeq 44$, $\chi^2_\pm \simeq 29$. The
 difference $H_i-P_i$ between the experimental and theoretic spectra are shown
 in Fig.\ref{dplexte0d}~and~\ref{dplextecd}. The difference for the 
 $\pi^+\pi^-$ distribution has a systematic spread of about  1\%. This spread
 was added to uncertainty of the theoretical $P_i^{(0)}$ distribution and the
 fit was redone (Table.\ref{tab3}). For all variants of fitting the values of
 $\rho$ meson widths $\Gamma_{\rho^0}$ and $\Gamma_{\rho^\pm}$ are consistent
 with relation (\ref{shir}), and fits were performed again under assumption
 that $g_{\rho^0\pi\pi}=g_{\rho^\pm\pi\pi}$ (Table.\ref{tab4})
 ($\Gamma_{\rho^0}$ was free parameter and  $\Gamma_{\rho^\pm}$ was calculated
 using expression (\ref{shir})). The $\rho^0$ and $\rho^\pm$ mass difference
 is consistent with zero, and the fits were redone again under assumption
 $m_{\rho^\pm}=m_{\rho^0}$ (Table \ref{tab5}). The experimental spectra
 fitted with theoretical distributions (variant 3 in Table \ref{tab5}) are
 shown in Fig.\ref{dplspe1} and \ref{dplspe2}.

 The parameters determined by fit are subject of various systematic errors
 originated from model dependence, from inaccuracy in background subtraction
 and in detection efficiency determination. To estimate errors due to
 uncertainty in background subtraction the experimental distributions were
 approximated with the sum of three histograms corresponding to the theoretical
 spectrum for the effect, $K_SK_L$ background and expected summary background
 from sources other then $K_SK_L$. The background events fractions were taken
 as free parameters of the fit. The values of these fractions obtained from
 the fit differs by less than 20\% from those estimated using equation
 (\ref{bg}).

 To study systematics related to inaccuracy in the extra photons simulation
 the corresponding correction factors were approximated by assuming the higher
 order polynomial, instead of linear function, to fit the dependence of these
 factors on the dipion mass shown in Fig.\ref{dplpop}. We also had tried the
 fit with linear approximation for the correction factors but with $a_0$ and
 $a_\pm$ as free parameters of the fit. The results of the fit
 $a_0=(-0.13\pm 0.02) \cdot 10^{-3}$ and $a_\pm=(0.12\pm0.2)\cdot 10^{-3}$ are
 consistent with those described above.

 The model dependence was estimated by using results of the fits performed
 under different assumptions  (Tables \ref{tab3}, \ref{tab4}, \ref{tab5}).
 
\section{Discussion}

 The SND results in comparison with other experimental data are shown 
 in Table \ref{tab6}.

 The fit results revealed  that the experimental data can be described by
 using only $\rho\pi$ intermediate state. The value of additional, other than
 $\rho\pi$, contribution $a_{3\pi}$ is consistent with zero (variant 2, in
 Table \ref{tab5}):
$$
 a_{3\pi} = (0.01 \pm 0.23 \pm 0.25) \times 10^{-5}  \mbox{~~MeV}^{-2},
$$
 Here the systematic error is related to uncertainties in background
 subtraction and detection efficiency determination. This result agrees with
 $a_{3\pi} = (0.08\pm0.45\pm0.37)\times 10^{-5}  \mbox{~~MeV}^{-2}$ reported
 in Ref.\cite{cmd2}. Using the $a_{3\pi}$ value and results of the
 $e^+e^-\to\pi^+\pi^-\pi^0$ cross section study \cite{phi98} the upper limit
 on the branching ratio of $\phi\to\pi^+\pi^-\pi^0$ decay without the $\rho\pi$
 intermediate state can be obtained:
 $$B(\phi\to\pi^+\pi^-\pi^0) < 6 \cdot 10^{-4} \mbox{~~(90\% CL)}$$
 This upper limit implies
 $$0.91 < { {\sigma_{3\pi}} \over {\sigma_{\rho\pi}} } < 1.09
 \mbox{~~(90\% CL)}$$
 Here $\sigma_{3\pi}$ is the total $e^+e^-\to\pi^+\pi^-\pi^0$ cross section,
 $\sigma_{\rho\pi}$ is  $e^+e^-\to\rho\pi\to\pi^+\pi^-\pi^0$ cross section.
 The upper and lower limits above correspond to the constructive and
 destructive interference between $a_{3\pi}$ and $\rho\pi$ related amplitudes.
 It may happen that $e^+e^- \to\phi\to\rho^\prime\pi$ amplitude is suppressed
 compared to $e^+e^- \to\omega(\omega^\prime)\to\rho^\prime\pi$ amplitude. In
 this case one expects $\mbox{Im}(a_{3\pi}) > \mbox{Re}(a_{3\pi})$. So we had
 performed also the fit with $\mbox{Im}(a_{3\pi})$ as free parameter and
 supposing $\mbox{Re}(a_{3\pi})=0$. The result of this fit also is consistent
 with zero: $$a_{3\pi} = i \cdot (-0.12 \pm 0.13 \pm 0.25)$$

 The fit with model taking into account $\rho-\omega$ mixing (variant 3 in
 Table \ref{tab5}) gave the results
$$
  K_{\omega\pi/\rho\pi} = 0.20 \pm 0.10 \pm 0.05 \mbox{~~}
  \phi_{\omega\pi/\rho\pi} = -125^\circ \pm 28^\circ.
$$
 The systematic error is related to uncertenity in background subtraction.
 The values obtained are consistent to the above given estimation
 $K_{\omega\pi/\rho\pi} \simeq 0.3$,
 $\phi_{\omega\pi/\rho\pi} \simeq -110^\circ$. The measured value of 
 $K_{\omega\pi/\rho\pi}$ deviates from zero by about two standard deviations.

 Parameter $s_1$ is related to the final state interaction of the $\rho$ and
 $\pi$ mesons and was measured to be (variant 5 in Table \ref{tab5}):
$$
 s_1 = 0.3 \pm 0.3 \pm 0.3
$$
 The main cause of the systematic error is again the background subtraction 
 uncertenity. The result is consistent with zero, but also does not contradict
 to the prediction of Ref.\cite{akfaz} ($s_1=1$ with about 10-20\% accuracy). 

 The measured neutral and charged $\rho$-mesons masses and their difference 
 are (variant 1 in Table \ref{tab4}):
$$
 m_{\rho^0} = 775.8 \pm 0.9 \pm 2.0 \mbox{~~MeV,~}
 m_{\rho^\pm} = 774.5 \pm 0.7 \pm 1.5 \mbox{~~MeV,~}
$$
$$
 m_{\rho^\pm}-m_{\rho^0} = -1.3 \pm 1.1 \pm 2.0 \mbox{~~MeV}.
$$
 The systematic error of $m_{\rho^\pm}$ is related to model dependence and 
 uncertainties in detection efficiency determination. Background subtraction
 inaccuracy also contributes in the systematic errors of $m_{\rho^0}$ and
 $m_{\rho^\pm}-m_{\rho^0}$. The results obtained are consistent with 
 $m_{\rho^\pm}=m_{\rho^0}$ as well as with the world average for the mass
 difference $-0.4 \pm 0.8$ MeV \cite{pdg}. The measured mass difference is also
 not in conflict with the prediction
 $m_{\rho^\pm}-m_{\rho^0} = -4.2 \pm 1.2$ MeV \cite{razmass}. If 
 $m_{\rho^\pm}=m_{\rho^0}$ is assumed in the fit, the $\rho$-meson mass turns
 out to be (variant 1 from Table.\ref{tab5}):
$$
 m_\rho=775.0 \pm 0.6 \pm 1.1 \mbox{~~MeV}
$$
 This value is consistent with the results of the $e^+e^-$ annihilation and
 $\tau$ decay experiments (Table \ref{tab6}). Note that world average for
 these experiments is $m_\rho=776 \pm 0.9$ MeV. But the PDG value \cite{pdg}
 $769.3 \pm 0.8$ MeV, which takes into account all experiments in which the
 $\rho$-meson mass was measured, contradicts to the reported SND result.

 Our results for the neutral and charged $\rho$-meson widths are (variant 1
 Table \ref{tab3}):
$$
 \Gamma_{\rho^0} = 151.1 \pm 2.6 \pm 3.0 \mbox{~~MeV,~}
 \Gamma_{\rho^\pm} = 149.9 \pm 2.3 \pm 2.0 \mbox{~~MeV}
$$
 The systematic error includes model dependence, uncertainty in background
 subtraction and in detection efficiency determination. Under assumption
 $g_{\rho^0\pi\pi}=g_{\rho^\pm\pi\pi}$ (variant 1 Table \ref{tab5}) the
 $\rho$-meson widths were found to be
$$ \Gamma_{\rho^0} = 149.8 \pm 2.2 \pm 2.0 \mbox{~~MeV,~}
   \Gamma_{\rho^\pm} = 150.9 \pm 2.2 \pm 2.0 \mbox{~~MeV}
$$
 These results are consistent with other experimental results (Table
 \ref{tab6}) as well as with the PDG world average $150.2 \pm 0.8$ MeV
 \cite{pdg}.

\section{Conclusions}

 In the SND experiment at VEPP-2M collider in Novosibirsk the dipion mass
 spectra were studied in the $e^+e^- \to \pi^+\pi^-\pi^0$ process at
 $\sqrt[]{s} = m_\phi$. The study is based on the data sample with about
 $0.5 \times 10^6$ experimental events. Spectra were analyzed within the VDM
 framework taking into account $e^+e^- \to \rho\pi$ transition, $\rho-\omega$
 mixing, final state interaction of the $\rho$ and $\pi$ mesons, possible
 transition $e^+e^- \to \pi^+\pi^-\pi^0$ through intermediate states different
 from $\rho\pi$ (for example, via $\rho^\prime\pi$). Within the limits of our
 accuracy, the experimental data can be described with the
 $e^+e^- \to \rho\pi \to \pi^+\pi^-\pi^0$ transition only. The upper limit on
 the $\phi\to 3\pi$ decay through intermediate states besides $\rho\pi$ is
 $B(\phi \to \pi^+\pi^-\pi^0)<6\cdot 10^{-4}$. In the $\pi^+\pi^-$ invariant
 mass spectrum the $\rho-\omega$ interference was observed at two standard
 deviations level. The measured mass and width values for the neutral and
 charged $\rho$-mesons are
 $m_{\rho^0} = 775.8 \pm 0.9 \pm 2.0$ MeV,
 $\Gamma_{\rho^0} = 151.1 \pm 2.6 \pm 3.0$ MeV,
 $m_{\rho^\pm} = 774.5 \pm 0.7 \pm 1.5$ MeV,
 $\Gamma_{\rho^\pm} = 149.9 \pm 2.3 \pm 2.0$ MeV.
 The difference between masses of the neutral and charged $\rho$-mesons was
 found to be $m_{\rho^\pm}-m_{\rho^0} = -1.3 \pm 1.1 \pm 2.0$ MeV.
 Under assumption $m_{\rho^\pm}=m_{\rho^0}$ the following $\rho$-meson mass
 value was obtained $m_\rho = 775.0 \pm 0.6 \pm 1.1$ MeV. Assuming the coupling
 constants equality $g_{\rho^0\pi\pi}=g_{\rho^\pm\pi\pi}$, the neutral and
 charged $\rho$-meson widths turn out to be
 $\Gamma_{\rho^0} = 149.8 \pm 2.2 \pm 2.0$ MeV and
 $\Gamma_{\rho^\pm} = 150.9 \pm 2.2 \pm 2.0$ MeV.

\section*{acknowledgments}

 The authors are grateful to N.N.Achasov and A.A.Kozhevnikov for useful
 discussions.

\newpage

\begin{table}
\begin{tabular}{cccccccc}
\hline
 $\sqrt[]{s}$ & $N_{3\pi}$ & $N_{bkg}$ & $N_{K_SK_L}$ & $N_{K^+K^-}$ &
 $N_{\eta\gamma}$ & $N_{\omega\pi}$ & $N_{2e2\gamma}$  \\
  (MeV) & & & & & & &  \\ \hline
 $\sim$ 1017 & 45913 & 410 & 331 & 3 & 42 & 15 & 19 \\
 $\sim$ 1018 & 101651 & 1067 & 901 & 8 & 104 & 24 & 29 \\
 $\sim$ 1019 & 137432 & 1696&1465&13&161&26&31 \\
 $\sim$ 1020 & 139319 & 1871&1674&12&147&27&31 \\
 $\sim$ 1021 & 68074 & 1024&903&7&78&17&20 \\
             & 492389 & 6088 &5274&43&532&109&130  \\ \hline 
\end{tabular}
\caption{Event numbers for the effect and estimated background at various
         energies. $N_{3\pi}$ - number of selected
	 $e^+e^- \to \pi^+\pi^-\pi^0$ events after background subtraction.
	 $N_{bkg}$ - the total estimated background. Contributions of separate
	 background sources are also shown. The last line contains the total
	 events numbers for all energy points.}
\label{tab1}
\end{table} 

\begin{table}
\begin{tabular}[t]{cccccccccccc}
$i$&$m_i$&$b^{(0)}_{i,i-4}$&$b^{(0)}_{i,i-3}$&$b^{(0)}_{i,i-2}$&
           $b^{(0)}_{i,i-1}$&$b^{(0)}_{i,i}$&$b^{(0)}_{i,i+1}$&
	   $b^{(0)}_{i,i+2}$&$b^{(0)}_{i,i+3}$&
	   $\delta P^{(0)}_i$&$H^{(0)}_i \cdot 10^4$\\
  &(MeV)&&&&&&&&&& \\ \hline
1&290&--&--&--&--&0.001&0.002&--&--&1.055&$0.10\pm0.04$ \\
2&310&--&--&--&--&0.015&0.009&0.001&--&1.004&$3.0\pm0.3$ \\
3&330&--&--&--&0.009&0.210&0.035&0.001&--&0.018&$49.6\pm1.1$ \\
4&350&--&0.001&0.002&0.026&0.295&0.045&0.002&0.001&0.015&$103.6\pm1.5$ \\
5&370&--&0.001&0.001&0.028&0.302&0.051&0.003&0.001&0.014&$154.0\pm1.9$ \\
6&390&0.001&0.001&0.003&0.034&0.317&0.058&0.003&0.001&0.013&$212.5\pm2.3$ \\
7&410&0.001&0.001&0.002&0.036&0.316&0.061&0.003&0.001&0.012&$267.9\pm2.5$ \\
8&430&--&--&0.001&0.040&0.313&0.065&0.004&0.001&0.012&$322.9\pm2.8$ \\
9&450&--&--&0.002&0.047&0.304&0.070&0.004&0.001&0.012&$378.2\pm3.0$ \\
10&470&--&--&0.002&0.050&0.292&0.071&0.005&0.001&0.012&$434.7\pm3.2$ \\
11&490&--&0.001&0.003&0.058&0.291&0.075&0.005&0.001&0.011&$485.9\pm3.4$ \\
12&510&--&0.001&0.003&0.061&0.273&0.073&0.006&0.001&0.011&$539.0\pm3.6$ \\
13&530&--&--&0.004&0.065&0.271&0.076&0.006&0.001&0.011&$578.3\pm3.8$ \\
14&550&--&0.001&0.005&0.067&0.258&0.073&0.006&0.001&0.011&$614.6\pm3.9$ \\
15&570&--&0.001&0.005&0.070&0.249&0.073&0.006&0.001&0.011&$645.6\pm4.0$ \\
16&590&--&0.001&0.006&0.072&0.239&0.071&0.006&0.001&0.011&$663.1\pm4.1$ \\
17&610&--&0.001&0.006&0.071&0.228&0.069&0.006&0.001&0.011&$668.3\pm4.1$ \\
18&630&--&0.001&0.006&0.070&0.222&0.067&0.006&0.001&0.011&$650.9\pm4.0$ \\
19&650&--&0.001&0.006&0.067&0.210&0.064&0.005&--&0.011&$628.8\pm4.0$ \\
20&670&--&0.001&0.006&0.065&0.204&0.059&0.004&--&0.011&$595.8\pm4.0$ \\
21&690&--&0.001&0.006&0.063&0.195&0.055&0.003&--&0.011&$544.5\pm3.8$ \\
22&710&--&0.001&0.005&0.060&0.187&0.053&0.003&--&0.012&$472.5\pm3.7$ \\
23&730&--&--&0.005&0.055&0.179&0.052&0.002&--&0.012&$380.7\pm3.5$ \\
24&750&--&--&0.003&0.050&0.169&0.049&0.002&--&0.013&$287.2\pm2.9$ \\
25&770&--&--&0.003&0.045&0.165&0.042&0.001&--&0.014&$181.0\pm2.2$ \\
26&790&--&--&0.002&0.035&0.154&0.034&--&--&0.017&$90.1\pm1.6$ \\
27&810&--&--&0.001&0.026&0.142&0.027&--&--&0.024&$32.3\pm0.9$ \\
28&830&--&--&--&0.021&0.113&0.034&--&--&0.039&$11.5\pm0.6$ \\
29&850&--&--&--&0.010&0.075&--&--&--&1.004&$3.4\pm0.4$ \\
30&870&--&--&--&0.001&0.045&--&--&--&1.023&$0.10\pm0.06$ \\
\end{tabular}
\caption{The normalized $\pi^+\pi^-$ invariant mass spectrum. $H^{(0)}_i$ -
         the relative fraction of events with $\pi^+\pi^-$ invariant mass
	 $m_i \pm 10$ MeV, efficiency matrix
	 $b^{(0)}_{ij}=a^{(0)}_{ij}\cdot \epsilon^{(0)}_{j}$ and
	 relative uncertainty $\delta P^{(\pm)}_i$ of the theoretical
	 $\pi^+\pi^-$ mass distribution are also given.}
\label{tab7}
\end{table}                     

\begin{table}
\begin{tabular}[t]{ccccccccccccc}
$i$&$m_i$&$b^{(\pm)}_{i,i-4}$&$b^{(\pm)}_{i,i-3}$&$b^{(\pm)}_{i,i-2}$&
           $b^{(\pm)}_{i,i-1}$&$b^{(\pm)}_{i,i}$&$b^{(\pm)}_{i,i+1}$&
	   $b^{(\pm)}_{i,i+2}$&$b^{(\pm)}_{i,i+3}$&$b^{(\pm)}_{i,i+4}$&
	   $\delta P^{(\pm)}_i$&$H^{(\pm)}_i \cdot 10^4$\\
  &(MeV)&&&&&&&&&&& \\ \hline
 1&290&--&--&--&--&0.030&0.006&--&--&-- &  1.004& $2.1\pm0.3$\\
 2&310&--&--&--&0.004&0.081&0.019&--&--&-- & 1.001&$12.2\pm0.7$ \\
 3&330&--&--&--&0.015&0.119&0.033&0.001&--&-- &  0.021&$33.8\pm1.1$ \\
 4&350&--&0.001&0.001&0.020&0.143&0.046&0.002&--&-- &0.016&$62.0\pm1.3$ \\
 5&370&--&--&0.001&0.033&0.162&0.059&0.004&--&-- &  0.013&$102.0\pm1.7$ \\
 6&390&0.001&--&0.002&0.039&0.171&0.064&0.005&--&-- & 0.011&$144.9\pm1.9$\\
 7&410&--&--&0.002&0.048&0.177&0.071&0.006&--&-- & 0.010&$194.4\pm2.3$\\
 8&430&--&--&0.002&0.056&0.183&0.077&0.008&0.001&-- & 0.009&$244.5\pm2.6$\\
 9&450&--&--&0.004&0.062&0.185&0.079&0.010&0.001&-- &  0.008&$298.5\pm2.8$\\
 10&470&--&--&0.005&0.067&0.190&0.085&0.011&0.001&-- & 0.007&$355.6\pm3.0$\\
 11&490&--&0.001&0.007&0.068&0.185&0.087&0.013&0.001&0.001 & 0.007&$417.9\pm3.3$\\
 12&510&--&--&0.008&0.074&0.189&0.088&0.014&0.002&0.001 & 0.006&$474.7\pm3.5$\\
 13&530&--&--&0.010&0.075&0.186&0.091&0.017&0.003&0.001 & 0.006&$526.1\pm3.7$\\
 14&550&--&0.001&0.010&0.076&0.183&0.090&0.017&0.003&0.001 & 0.006&$578.1\pm3.9$\\
 15&570&--&0.001&0.013&0.078&0.177&0.093&0.017&0.003&0.001 & 0.006&$620.4\pm4.0$\\
 16&590&--&0.001&0.013&0.081&0.174&0.094&0.019&0.004&0.001 & 0.006&$645.8\pm4.1$\\
 17&610&--&0.002&0.015&0.079&0.170&0.094&0.019&0.004&0.001 & 0.006&$667.3\pm4.1$\\
 18&630&--&0.002&0.015&0.079&0.168&0.091&0.020&0.005&0.001 & 0.006&$676.0\pm4.1$\\
 19&650&0.001&0.003&0.016&0.080&0.166&0.090&0.020&0.005&0.001 & 0.006&$673.4\pm4.1$\\
 20&670&0.001&0.003&0.016&0.078&0.163&0.089&0.021&0.004&0.001 & 0.006&$660.6\pm4.1$\\
 21&690&0.001&0.003&0.017&0.078&0.163&0.090&0.023&0.005&0.001 & 0.006&$631.5\pm4.0$\\
 22&710&0.001&0.003&0.016&0.078&0.162&0.093&0.021&0.005&0.001 & 0.006&$570.7\pm3.8$\\
 23&730&0.001&0.004&0.017&0.078&0.163&0.090&0.022&0.005&0.001 & 0.007&$498.4\pm3.5$\\
 24&750&0.001&0.004&0.017&0.077&0.166&0.094&0.023&0.004&0.001 & 0.008&$381.2\pm3.0$\\
 25&770&0.001&0.003&0.017&0.077&0.160&0.091&0.020&0.002&-- & 0.010&$259.5\pm2.5$\\
 26&790&0.001&0.004&0.017&0.076&0.171&0.096&0.018&0.002&-- & 0.012&$147.6\pm1.9$\\
 27&810&0.001&0.003&0.016&0.073&0.167&0.087&0.014&--&-- & 0.018&$72.6\pm1.3$\\
 28&830&0.001&0.003&0.013&0.070&0.170&0.054&--&--&-- & 0.029&$30.8\pm0.9$\\
 29&850&0.001&0.002&0.013&0.060&0.171&0.029&--&--&-- & 1.001&$11.7\pm0.5$\\
 30&870&0.001&0.001&0.012&0.045&0.057&--&--&--&-- & 1.005&$5.6\pm0.4$\\
\end{tabular}
\caption{The normalized $\pi^\pm\pi^0$ invariant mass spectrum. $H^{(\pm)}_i$ -
         the relative fraction of events with $\pi^\pm\pi^0$ invariant mass
	 $m_i \pm 10$ MeV, efficiency matrix
	 $b^{(\pm)}_{ij}=a^{(\pm)}_{ij}\cdot \epsilon^{(\pm)}_{j}$ and
	 relative uncertainty $\delta P^{(\pm)}_i$ of the theoretical
	 $\pi^\pm\pi^0$ mass distribution are also given.}
\label{tab8}
\end{table}

\begin{table}
\begin{tabular}[t]{c|cccc}
 $N$&1&2&3&4 \\
 $m_{\rho^0}-770$ (MeV) & 6.1$\pm$1.0&6.1$\pm$1.0&4.9$\pm$1.4&7.8$\pm$1.6 \\
 $\Gamma_{\rho^0}$(MeV)&151.1$\pm$2.6&150.7$\pm$3.5&156.0$\pm$4.0&151.2$\pm$2.6 \\
 $m_{\rho^\pm}-m_{\rho^0}$ (MeV)&-1.7$\pm$1.2&-1.8$\pm$1.2&-0.6$\pm$1.6&-1.9$\pm$1.2 \\
 $\Gamma_{\rho^\pm}$(MeV)&149.9$\pm$2.3&149.5$\pm$3.4&149.4$\pm$2.3&150.2$\pm$2.3 \\
 $\mbox{Re}(a_{3\pi}) \cdot 10^5$(MeV)$^{-2}$&0&0.04$\pm$0.25&0&0 \\
 $K_{\omega\pi/\rho\pi}$ &0&0&0.3$\pm$0.1&0 \\
 $\phi_{\omega\pi/\rho\pi}$ &0&0&-90.0$\pm$18.0&0 \\
 $s_1$ &0&0&0&0.3$\pm$0.3 \\
 $\chi^2_0$&27.1&26.9&19.0&25.0 \\
 $\chi^2_\pm$&28.9&29.1&29.5&29.6 \\
 $\chi^2/N_{df}$ &$56.0/54$&$56.0/53$&$48.5/52$&$54.6/53$ \\
\end{tabular}
\caption{Fit results for the $\pi^+\pi^-$ and $\pi^\pm\pi^0$ mass spectra.}
\label{tab3}
\end{table}
\begin{table}
\begin{tabular}[t]{c|cccc}
 $N$&1&2&3&4 \\
 $m_{\rho^0}-770$ (MeV) &5.8$\pm$0.9&5.8$\pm$1.0&4.3$\pm$1.3&7.3$\pm$1.3 \\
 $\Gamma_{\rho^0}$(MeV)&149.9$\pm$2.2&149.5$\pm$3.1&149.3$\pm$2.3&149.9$\pm$2.2 \\
 $m_{\rho^\pm}-m_{\rho^0}$(MeV)&-1.3$\pm$1.1&-1.3$\pm$1.1&0.3$\pm$1.5&-1.4$\pm$ 1.1 \\
 $\mbox{Re}(a_{3\pi}) \cdot 10^5$(MeV)$^{-2}$&0&0.04$\pm$0.24&0&0 \\
 $K_{\omega\pi/\rho\pi}$ &0&0&0.2$\pm$0.1&0 \\
 $\phi_{\omega\pi/\rho\pi}$ &0&0&-125.0$\pm$26.0&0 \\
 $s_1$ &0&0&0&0.3$\pm$0.3 \\
 $\chi^2_0$&26.9&26.8&24.0&24.9 \\
 $\chi^2_\pm$&29.9&30.0&30.0&30.7 \\
 $\chi^2/N_{df}$ &$56.8/55$&$56.8/54$&$54.3/53$&$55.6/54$ \\
\end{tabular}
\caption{Fit results for the $\pi^+\pi^-$ and $\pi^\pm\pi^0$ mass spectra
         under assumption $g_{\rho^0\pi\pi} = g_{\rho^\pm\pi\pi}$.}
\label{tab4}
\end{table}
\begin{table}
\begin{tabular}[t]{c|cccc}
 $N$&1&2&3&4 \\
 $m_{\rho^0}-770$ (MeV) &5.0$\pm$0.6&5.0$\pm$0.7&4.5$\pm$0.7&6.4$\pm$1.4 \\
 $\Gamma_{\rho^0}$(MeV)&149.8$\pm$2.2&149.7$\pm$3.1&149.3$\pm$2.3&149.9$\pm$2.2 \\
 $\mbox{Re}(a_{3\pi}) \cdot 10^5$(MeV)$^{-2}$&0&0.01$\pm$0.23&0&0 \\
 $K_{\omega\pi/\rho\pi}$ &0&0&0.2$\pm$0.1&0 \\
 $\phi_{\omega\pi/\rho\pi}$ &0&0&-125.0$\pm$28.0&0 \\
 $s_1$ &0&0&0&0.3$\pm$0.3 \\
 $\chi^2_0$&28.2&28.0&24.0&26.2 \\
 $\chi^2_\pm$&30.0&30.0&30.0&30.8 \\
 $\chi^2/N_{df}$ &$58.2/56$&$58.0/55$&$58.0/54$&$57.1/55$ \\
\end{tabular}
\caption{Fit results for the $\pi^+\pi^-$ and $\pi^\pm\pi^0$ mass spectra,
         under assumptions $g_{\rho^0\pi\pi} = g_{\rho^\pm\pi\pi}$ and
	 $m_{\rho^0}=m_{\rho^\pm}$.}
\label{tab5}
\end{table}

\begin{table}
\begin{tabular}[t]{l|lll}
 &$m_{\rho^0}$(MeV)&$m_{\rho^\pm}$(MeV)&$m_{\rho^\pm}-m_{\rho^0}$(MeV) \\
 SND&$775.8 \pm 0.9 \pm 2.0$&$774.5 \pm 0.7 \pm 1.5$&$-1.3 \pm 1.1 \pm 2.0$ \\
 SND$^{\star}$&$775.0 \pm 0.6 \pm 1.1$&$775.0 \pm 0.6 \pm 1.1$& \\
 CMD-2 \cite{kmd}&$775.3 \pm 0.6 \pm 0.2$&& \\
 ALEPH \cite{aleph}&&$776.4 \pm 0.9 \pm 1.5$&$0.0 \pm 1.0$ \\
 CLEO \cite{cleo}&&$774.9 \pm 0.5 \pm 0.9$& \\
 CBAR \cite{cbar}&$762.3 \pm 0.5 \pm 1.2$&$763.0 \pm 0.3 \pm 1.3$&$-1.6 \pm 0.6 \pm 1.7$ \\
 PDG-1\cite{pdg}&$776.0 \pm 0.9$&$776.0 \pm 0.9$& \\
 PDG-2 \cite{pdg}&$769.3 \pm 0.8$&$769.3 \pm 0.8$&$-0.4 \pm 0.8$ \\
\hline
 &$\Gamma_{\rho^0}$(MeV)&$\Gamma_{\rho^\pm}$(MeV)&$a_{3\pi}\times 10^5$(MeV)$^{-2}$ \\
SND&$151.1 \pm 2.6 \pm 3.0$&$149.9 \pm 2.3 \pm 2.0$& \\
SND$^{\star\star}$&$149.8 \pm 2.2 \pm 2.0$&$150.9 \pm 2.2 \pm 2.0$&$0.01 \pm 0.23 \pm 0.25$ \\
CMD-2 \cite{kmd,cmd2}&$147.7 \pm 1.3 \pm 0.4$&&$0.08 \pm 0.45 \pm 0.37$ \\
ALEPH \cite{aleph}&&$150.5 \pm 1.6 \pm 6.3$& \\
CLEO \cite{cleo}&&$149.0 \pm 1.1 \pm 0.7$& \\
CBAR \cite{cbar}&$147.0 \pm 2.5$&$149.5 \pm 1.3$&\\
PDG-1\cite{pdg}&$150.5 \pm 2.7$&$150.5 \pm 2.7$& \\
PDG-2\cite{pdg}&$150.2 \pm 0.8$&$150.2 \pm 0.8$& \\
\hline
  &$K_{\omega\pi/\rho\pi}$&$\phi_{\omega\pi/\rho\pi}$&$s_1$ \\
SND&$0.20 \pm 0.10 \pm 0.05$&$-125^\circ \pm 28^\circ$&$0.3 \pm 0.3 \pm 0.3$ \\
\end{tabular}
\caption{Comparison of the results of various experiments. SND -- this work;
         SND$^\star$ -- the $\rho$ meson mass from our fit under
	 $m_{\rho^0}=m_{\rho^\pm}$ assumption; SND$^{\star\star}$ -- the
	 $\rho^0$ and $\rho^\pm$ widths from our fit under
	 $g_{\rho^0\pi\pi} = g_{\rho^\pm\pi\pi}$ assumption; PDG-1 -- the
	 world average values of $e^+e^-$ annihilation and $\tau$ decay
	 experiments only; PDG-2 -- the world average values for all
	 experiments.}
\label{tab6}
\end{table}

\vspace{-2cm}
\vspace{-2cm}

\vspace{-2cm}
\vspace{-2cm}

\vspace{-2cm}
\vspace{-2cm}

\end{document}